\documentclass[%
 reprint,
nofootinbib,
 amsmath,amssymb,
 aps,
]{revtex4-2}

\usepackage{graphicx}
\usepackage{dcolumn}
\usepackage{bm}

\usepackage{physics}
\usepackage{standalone}
\usepackage{tikz}
\usepackage[dvipsnames]{xcolor}
\usepackage[disable]{todonotes}
\usepackage{soul}

\definecolor{pcolor}{RGB}{255,10,255}

\definecolor{linkcol}{RGB}{18, 94, 173}     
\definecolor{citecol}{RGB}{166, 45, 124}    
\definecolor{urlcol}{RGB}{0, 128, 112}      

\usepackage[
    colorlinks=true,
    linkcolor=linkcol,
    citecolor=citecol,
    urlcolor=urlcol,
]{hyperref}

\def\be{\begin{equation}}
\def\ee{\end{equation}}
\def\ba{\begin{align}}
\def\ea{\end{align}}
\def\bs{\begin{split}}
\def\es{\end{split}}
\def\bea{\begin{eqnarray}}
\def\eea{\end{eqnarray}}

\def\ha{\hat{a}}
\def\hb{\hat{b}}

\definecolor{darkred}{rgb}{0.8,0.1,0.1}

\newcommand{\upd}{\mathrm{d}}

\makeatletter
\if@todonotes@disabled

\else

\fi
\makeatother
\begin{document}

\preprint{}

\title{Wavelet Matrix Product States for Quantum Fields}

\author{Molly Kaplan}
\email{molly.kaplan@inria.fr}
\author{Antoine Tilloy}%
 \email{antoine.tilloy@mineparis.psl.eu}
\affiliation{Laboratoire de Physique de l’École Normale Supérieure, Mines Paris, Inria, CNRS, ENS-PSL, Centre Automatique et Systèmes (CAS),
Sorbonne Université, PSL Research University, Paris, France}%

%
\begin{abstract}
We introduce a variational method to solve continuum quantum models with discrete tensor network techniques. The method leverages wavelet matrix product states (wMPS): matrix product states built on top of sufficiently regular ($N\geq 6$) Daubechies scaling functions. These states live in the continuum field theory Fock space, have finite energy density, and can be optimized with standard algorithms, without restriction to free theories. Further, exploiting the multi-resolution analysis built into wavelets, and its quantum circuit description, we can iteratively refine wMPS to obtain accurate approximations at arbitrarily fine length-scales. We showcase the efficiency of the method on the Lieb-Liniger model, computing energy density and correlation functions.
\end{abstract}

\maketitle


\section{Introduction}
Tensor network states are a central tool in the quest to solve the quantum many-body problem classically. Tailored to the low energy states of local Hamiltonians, they already allow the simulation of large many-body systems numerically, with matrix product states (MPS)~\cite{White_1992,White_1993,Fannes_Nachtergaele_Werner_1992,schollwock2011review,Cirac_PerezGarcia_Schuch_Verstraete_2021,Orus_2014} in $1$ space dimension, and projected entangled-pair states (PEPS) \cite{verstraete2004renormalizationalgorithmsquantummanybody,Cirac_PerezGarcia_Schuch_Verstraete_2021,Corboz_2016,Vanderstraeten_Haegeman_Corboz_Verstraete_2016,Zheng_2017,Liao_Liu_Wang_Xiang_2019} in $2$ dimensions and more.

As their name suggests, tensor network methods are tied to a lattice. To study continuum models, like relativistic and non-relativistic quantum field theories (QFT), one has two strategies. First, one can discretize the model, solve it with standard tensor network techniques numerically, and then extrapolate the results to the continuum limit. In many cases, this is impressively accurate~\cite{haghshenas2021,Milsted_Haegeman_Osborne_2013,Banuls_Cichy_Cirac_Jansen_2013,Dempsey_Gluck_Pufu_Sogaard_2026} as one can use the broad range of optimization methods developed for the lattice. However, one loses the guarantees of a variational method: it is difficult to know for sure if results on a finer lattice are actually better than those on a coarser one. Alternatively, one can take the continuum limit of the ansatz itself, and apply it to the continuum model directly. This yields variational results that need not be extrapolated. In $1$ space dimension, continuous matrix product states (CMPS)~\cite{Verstraete_2010,PhysRevB.88.085118,ganahl2017}, and their relativistic refinement~\cite{PhysRevD.104.L091904,Tilloy_2021} work well, particularly in the thermodynamic limit. But they require new \emph{ad hoc} optimization methods, and not all of the tensor network toolbox has been translated to the continuum yet. In $2$ space dimensions and more, the situation is even more dire, and the continuous equivalent of PEPS~\cite{jennings2015,PhysRevX.9.021040} is not yet practically usable for numerics. 

In this article, we aim to develop a third way: use a discrete ansatz to solve a continuum theory variationally. To decompose states of quantum fields in position space, we consider an orthonormal family of sufficiently regular Daubechies wavelets~\cite{Daubechies1988OrthonormalBO, Daubechies_1992}, which are a powerful tool in signal processing and numerical analysis. These functions are arbitrarily refinable into a discrete basis of $L^2(\mathbb{R})$, differentiable, and compactly supported. Taking a subset of them, the Daubechies scaling functions at a fixed resolution $r$, one can build a Fock space, which is a subspace $\mathcal{H}^r$ of the continuum Hilbert space $\mathcal{H}$. This Fock space has a natural tensor product structure, which respects spatial locality. It is thus natural to write candidate low energy states within $\mathcal{H}^r$ as MPS, which is what we propose doing. 

These wavelet-MPS (wMPS) have appealing properties: they are compatible with standard MPS algorithms (and thus applicable to interacting models), remain fully variational, and can be iteratively fine grained. One can start with a variational approximation of the ground state at a coarse scale, obtained very cheaply, and then refine it into a better approximation at a scale twice as fine. In practice, we easily chain more than a dozen such successive refinements without numerical instabilities. Conceptually, the construction is not in any way tied to $1$ space dimension, and is transparently extended to $d\geq 2$ space dimensions, albeit with likely tremendous numerical costs.

There has already been considerable discussion of wavelets in QFT and tensor networks, which it is important to contrast with what we are proposing.

First, there is a natural connection between wavelets and tensor network states, via the multi-scale entanglement renormalization ansatz (MERA)~\cite{PhysRevLett.99.220405,Evenbly_Vidal_2014}. More precisely, it has been shown that wavelet transforms can be written as a quantum circuit corresponding to a free fermionic or bosonic MERA \cite{PhysRevLett.116.140403, PhysRevX.8.011003, PhysRevA.97.052314, Witteveen_Scholz_Swingle_Walter_2022, 10.21468/SciPostPhys.10.6.143}. Since the MERA itself is connected to the AdS/CFT correspondence, this has provided insights into holography~\cite{PhysRevD.86.065007, qi2013exactholographicmappingemergent, singh2016holographicconstructionquantumfield}. However, despite important analytical insights, this MERA-wavelet connection has been used numerically only for free field theories.

Second, wavelets have long been considered as a candidate to discretize quantum fields in a controllable way, both for analytical questions \cite{PhysRevD.87.116011} and for simulation on quantum computers~\cite{PhysRevA.92.032315}. Perhaps most closely related to our work is Alves~\cite{Alves_2024}, which expands quantum fields in position space using Daubechies scaling functions and then uses a wavelet MERA to approximate the ground state. As in previous work connecting wavelets and MERA, practical applications remain restricted to free field theories. Beyond free theories, Daubechies wavelets have recently been used to discretize a Hamiltonian formulation of 1+1 dimensional $\phi^4$ theory in momentum space~\cite{basak2026hamiltonianformulation11dimensionalphi4, basak2026novelhamiltonianformulation11}, and solve it with Hamiltonian truncation (\textit{i.e.} exact diagonalization on a subspace of the total Hilbert space). Because the Hamiltonian is non-local in momentum, this latter construction is not natural to use with tensor networks, and its cost remains exponential in system size.

Finally, there are recent discretizations of QFT that combine with tensor network states well, but that do not use wavelets. In~\cite{PhysRevLett.128.230401}, the authors cut space into small segments (which is akin to using the lowest order Daubechies wavelets, or Haar wavelets) on which they approximately diagonalize the local Hamiltonian. This provides a natural discrete tensor factorization of the Hilbert space, allowing one to find the ground state with the density matrix renormalization group (DMRG). The method is precise, but not strictly variational: connecting the segments continuously adds an infinite penalty term to the Hamiltonian, which needs to be approximated. Most recently, as we were putting the final touches to the present article, Shankar, Van Acoleyen, and Haegeman~\cite{shankar2026finiteelementmatrixproductstates} have proposed a finite element discretization that is strictly variational, and preserves the locality structure that tensor network states need. Their construction is quite different from and complementary to ours, with some benefits -- their family of functions has smaller support -- and drawbacks -- the finite elements are not orthonormal, which requires a modification of standard tensor network algorithms.

In Section~\ref{sec:QFTdecompose}, we present the mathematics of Daubechies scaling functions (the first half of Daubechies wavelet theory), before detailing our variational wavelet MPS approach. In Section~\ref{sec:wMPSfixed}, we use this fixed-resolution approach to study the ground state properties of the Lieb-Liniger model and assess convergence. In Section~\ref{sec:refinement}, we introduce Daubechies wavelet functions (the second half of Daubechies wavelet theory), which allow us to leverage the power of multi-resolution analysis, and iteratively refine our states until they are numerically free of finite scale artifacts.
\section{Discrete basis for quantum fields}\label{sec:QFTdecompose}
In this section, we present the continuous Fock space of non-relativistic QFT, using the Lieb-Liniger model as an example. We then give three criteria for a good discrete orthonormal family of functions, adapted to variational MPS, which ultimately motivate our choice of Daubechies scaling functions. We then review some of their key properties.

\subsection{The (non-relativistic) QFT Hilbert space}
To make our variational method concrete, we first give the prototypical example to which it applies: the Lieb-Liniger model. Our method may be used for other models with the same non-relativistic kinetic term and sufficiently local interactions, including models with more particle species, without difficulty. 

The Lieb-Liniger model is a theory of non-relativistic bosons on an interval $I$, interacting with a contact potential. Its Hamiltonian takes the following second-quantized form:
\begin{equation}\label{eq:LL}
    H=\int_I \partial_x \hat{\psi}^\dag \partial_x \hat\psi + c\, \hat\psi^\dag \hat\psi^\dag \hat\psi \hat\psi - \mu \, \hat\psi^\dag \hat\psi,
\end{equation}
where $c$ is the interaction strength, $\mu$ the chemical potential, and $\hat\psi(x)$ the bosonic annihilation operator, with commutation relations $[\hat\psi(x),\hat\psi(y)^\dag]=\delta(x-y)$ and $[\hat\psi(x),\hat\psi(y)]=[\hat\psi(x)^\dag,\hat\psi(y)^\dag]=0$. This model is integrable, which makes it a convenient benchmark: it is possible to compute energy density errors, and to cleanly separate finite scale from finite bond dimension effects by comparing with CMPS values. 

The Hilbert space of this model is the Fock space $\mathcal{F}(L^2(I))$, obtained by creating excitations on top of the free Fock vacuum $\ket{0}$ annihilated by all $\hat\psi(x)$. A generic state $\ket{\Psi}$ is a sum of states with fixed particle numbers
\begin{equation}
  \begin{split}
    \ket{\Psi}=& \displaystyle\sum\limits_{m\geq 0} \int_{I^m} \upd x_1...\upd x_m \; \varphi_m(x_1,...,x_m) \\
    &\quad \quad\quad \quad \times\, \hat\psi^\dag (x_1) ... \hat\psi^\dag (x_m) \ket{0},\label{eq:secondq}
  \end{split}
\end{equation}
where $\varphi_m$ is the $m$-particle wavefunction. Our goal is to write a discretely parameterized MPS that belongs to this Fock space, and that approximates the true ground state $\ket{\text{GS}}$ of $H$ well, especially in the thermodynamic limit where $I\rightarrow \mathbb{R}$.

\subsection{Requirements for a discrete orthonormal family}
To write an MPS in the QFT Hilbert space, we first need a discrete orthonormal family of states, with a simple tensor product structure, spanning a subspace $\mathcal{H}^r$ of $\mathcal{F}(L^2(\mathbb{R}))$ [where $r$ is a \emph{fixed} scale, associated to a typical width $\Delta x = 1/2^r$]. A natural approach is to take first a discrete orthonormal family of functions $f_n^r \in L^2(\mathbb{R})$, and then consider the tensor product of the Fock spaces associated to each mode function $f_n^r$: $\mathcal{H}^r = \bigotimes_n \mathcal{F}(f_n^r)$. This is the approach we follow in this article. The question now becomes: what makes a good orthonormal family of $L^2(\mathbb{R})$ for our purposes?

For our method to be ultimately efficient, we require three properties:
\begin{enumerate}
  \item \textbf{Locality} -- The Hamiltonian $H$ is local, and we would like it to remain local once written in our discrete tensor product subspace. This will ensure that $H$ can be written exactly in MPO form, that its ground state has area law entanglement (up to log corrections) for this tensor product structure, and thus that MPS approximate its ground state well. We thus demand that the discrete orthonormal family of functions we consider is \emph{compactly supported}.
  \item \textbf{Regularity} -- The kinetic term of a non-relativistic QFT Hamiltonian, as in \eqref{eq:LL}, involves the first derivative of the field operator. Thus, to obtain finite kinetic energies, we demand that the functions we consider are \emph{differentiable}. This is crucial to keep our method truly variational.
  \item \textbf{Inclusion} -- Refining in scale, \textit{i.e.} using functions of smaller width, should allow us to systematically get better variational approximations. To get this property, we demand that the vector space spanned by functions at resolution $r$ is \emph{included} in the vector space spanned by functions at resolution $r+1$: $\text{span}\{f_n^r,\; n\} \subset \text{span}\{f_n^{r+1},\; n\} $. Indeed, if this is the case, the inclusion can be promoted to Fock spaces, and implies that the Hilbert space considered grows with $r$: $\mathcal{H}^{r}\subset\mathcal{H}^{r+1}$.
\end{enumerate}
Daubechies scaling functions with parameter $N\geq 6$ satisfy all the properties listed above. However, as a start, it is helpful to see why these conditions are non-trivial, and why other natural choices of functions, which have been considered previously, violate at least one of our requirements.

The simplest choice is \emph{naive discretization}, where discrete ladder operators at position $n$ and resolution $r$ are defined in terms of the bosonic field operators as
\begin{equation}
    \hat a_n^r = 2^{r/2} \int_{2^{-r}n}^{2^{-r}(n+1)} \upd x\,  \hat\psi(x)  = \int_{\mathbb{R}} \upd x\, s_n^r(x) \, \hat\psi(x),
\end{equation}
where we have introduced the so-called Haar scaling functions (which are simply windows) 
\begin{equation}
    s_n^r(x) = \begin{cases}
        2^{r/2}, & \text{if } 2^{-r}n \leq x < 2^{-r}(n+1), \\
        0, & \text{otherwise}.
    \end{cases}
\end{equation}
These functions indeed form an orthonormal family at fixed $r$, have finite support, and the Haar scaling functions at scale $r$ are simple sums of Haar scaling functions at scale $r+1$. However, they are not differentiable. Thus, all the states in this Hilbert space but the trivial vacuum have infinite energy. The standard fix is to regulate the kinetic term by taking finite differences instead of derivatives, but this means changing $H$, and thus losing the strict variational nature of the approach.

At least if we consider a finite interval $I=[0,L]$, a natural choice would be a Fourier decomposition. As in \cite{schmoll2023hamiltoniantruncationtensornetworks}, one could expand the bosonic operator $\hat\psi(x)$ in quantized plane waves, 
\begin{equation}
    \hat \psi(x) = \sum_{k \in \mathbb{Z}}\frac{1}{\sqrt{L}} e^{2\pi i kx/L} \hat a_k .
\end{equation}
This generates wildly non-local interaction terms. For instance, 
\begin{equation}
  \int_{[0,L]} \hat\psi^\dag \hat\psi^\dag \hat\psi \hat\psi = \frac{1}{L} \sum_{k_1,k_2,k_3} \hat a_{k_1}^\dag \hat a_{k_2}^\dag \hat a_{k_3} \hat a_{k_1+k_2-k_3}.
\end{equation}
This non-locality is the sign that entanglement in momentum generically follows a volume law, and thus the corresponding tensor product structure is not adapted to MPS when $L\rightarrow +\infty$\footnote{In \cite{schmoll2023hamiltoniantruncationtensornetworks}, the authors consider a small coupling, expand around the free ground state, and thus keep a small (though still volume law) entanglement for finite $L$.}.

Finally, one could consider smooth localized bump functions, for example,
\begin{equation}
    f_n^r(x) = \begin{cases}
        \frac{2^{r/2}}{C} \exp\left(\frac{1}{4(2^rx -n)^2-1}\right), & \text{if } |x| < 1/2, \\
        0, & \text{otherwise},
    \end{cases}
\end{equation}
with $C=\sqrt{\int_{-1/2}^{1/2} \upd x \exp\left(2/(4x^2-1)\right)}$. These functions are local, differentiable, and orthonormal at fixed $r$. However, they violate our last inclusion condition: a bump function at resolution $r$ cannot be written as a linear combination of bump functions at resolution $r+1$.

\subsection{Daubechies scaling functions}\label{sec:Daubechies}

Daubechies scaling functions verify the three conditions we demand, on top of other favorable properties. They are presented thoroughly in standard references~\cite{Daubechies_1992, Mallat_2009}. In what follows, we introduce them in a slightly different way, to show why they are natural for our purposes.

The starting point is a single function $s$, which we assume supported on $[0,N-1]$ with $N$ even. We will specify it only later, once we understand what it should verify. To construct our family of functions from this single bump (or scaling) function, we introduce the dyadic dilation operator $\mathcal{D}$ and translation operator $\mathcal{T}$ defined in the following way: 
\begin{equation}
    \begin{split}
        \mathcal{D}^r s (x)=& 2^{r/2} s(2^r x) \text{ for } r \in \mathbb{Z} \\
        \mathcal{T}^n s(x) =& s(x-n) \text{ for } n \in \mathbb{Z}.
    \end{split}
\end{equation}
Combining these operations, we write the $n$ translated scaling functions at resolution $r$ as
\begin{equation}
  s_n^r(x) := \mathcal{D}^r  \mathcal{T}^n s(x) = 2^{r/2}s(2^r x - n).
\end{equation}
We now demand that the functions $s_n^r(x)$ are orthonormal at fixed resolution $r$, and write $\mathcal{S}_r= \text{span}\{s_n^r(x)\} |_{n \in \mathbb{Z}}$. Then $\mathcal{S}_r$ is the subspace of $L^2(\mathbb{R})$ that the $s_n^r(x)$ span. 

For our inclusion condition to be verified, we further ask that $s_n^r$ can be written as a linear combination of $s_m^{r+1}$. Since the functions are compactly supported, this has to be a finite sum, called a \emph{refinement relation}:
\begin{equation}\label{eq:WTs}
    s_n^r(x) = \sum_{i=0}^{N-1}h_{i} s_{2n+i}^{r+1}(x),
\end{equation}
where $h_i$ are the \emph{weights} or filter coefficients. This relation naturally implies that
\begin{equation}\label{eq:sub_inc}
  \mathcal{S}_0 \subset \mathcal{S}_1 \subset \dots \subset \mathcal{S}_r \subset \mathcal{S}_{r+1} \subset \dots \subset L^2(\mathbb{R}) \, .
\end{equation}
For \eqref{eq:WTs} to be true, the bump function $s$ has to be very special. Indeed, using the definition of the dyadic dilation operator, it is equivalent to the fixed-point equation
\begin{equation}\label{eq:RGs}
  s(x) = \sum_{i=0}^{N-1} h_i s_i^1(x) = \underset{\mathfrak{h}\cdot s(x)}{\underbrace{\sqrt{2} \sum_{i=0}^{N-1} h_i s(2x-i)}}\, .
\end{equation}
This is an infinite dimensional linear equation of the form $s= \mathfrak{h} \cdot s$, and thus it has a non-trivial solution only if $\mathfrak{h}$ has an eigenvalue $1$.

Demanding that the fixed-point equation has a non-trivial solution, that the resulting solution is regular, and that the integer-translated solutions are orthogonal is not quite sufficient to fix the weights $h_i$. Daubechies scaling functions are obtained uniquely by demanding on top that one can represent polynomials of degree $N/2-1$ exactly\footnote{Note that Daubechies scaling functions are \emph{not} the most regular scaling functions at a fixed support length $N-1$. However, they become differentiable at $N=6$ and, to our knowledge, there exists no other orthonormal family that is both differentiable on a smaller support and verifies our conditions. Hence, for our purposes, the expressiveness condition \eqref{eq:polys} is a very good proxy for regularity, even if it is not equivalent in general.}, \textit{i.e.} that there exist coefficients $c_n^{(k)}$ such that
\begin{equation}\label{eq:polys}
  x^k = \sum_{n=-\infty}^{\infty} c^{(k)}_n s_n(x),\text{ for }0 \leq k \leq N/2-1 \, .
\end{equation}
This property implies that the regularity of $s$ increases with $N$, and for $N \geq 6$ the function $s$ has a continuous first derivative \cite{Daubechies_Lagarias_1991, Daubechies_Lagarias_1992}. The weights $h_i$ are obtained by solving the non-linear algebraic equations implied by the orthonormality constraint, the fixed-point equation \eqref{eq:RGs}, and the representation equation \eqref{eq:polys}. These equations can be solved exactly with radicals up to $N=10$ (and otherwise computed numerically). 

Once the weights are known, one can solve eq.~\eqref{eq:RGs} numerically to find $s$. One simple strategy is to write eq.~\eqref{eq:RGs} for $x$ set to integer $n$, which gives a finite linear system since $s$ is compactly supported. Once $s(n)$ is known, one can plug the values in the right-hand side of  \eqref{eq:RGs} to get $s$ at half integer values. Recursively, one gets exact values for all dyadic rationals, which are dense in $\mathbb{R}$. As this fractal construction may suggest, the resulting function $s$ takes a peculiar form, at least for $N>2$ ($N=2$ corresponds to Haar scaling functions). At $N=6$, we can visualize the refinement relation \eqref{eq:WTs}
\begin{widetext}
\begin{center}
    
\begin{tikzpicture}

\node at (2.47,-2.5) {$+$};
\node at (5,-2.5) {\includegraphics[scale=0.3]{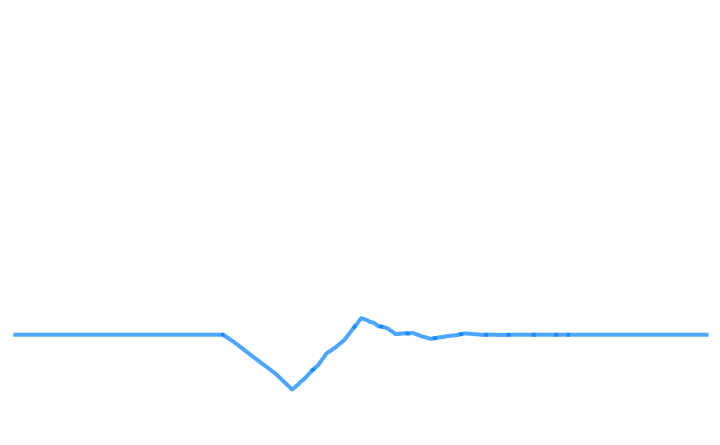}};
\node at (7.35,-2.5) {$+$};
\node at (9.7,-2.5) {\includegraphics[scale=0.3]{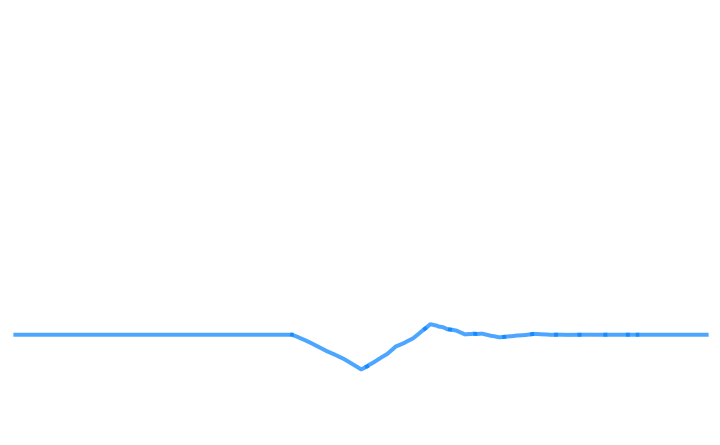}};
\node at (12.05,-2.5) {$+$};
\node at (14.4,-2.5) {\includegraphics[scale=0.3]{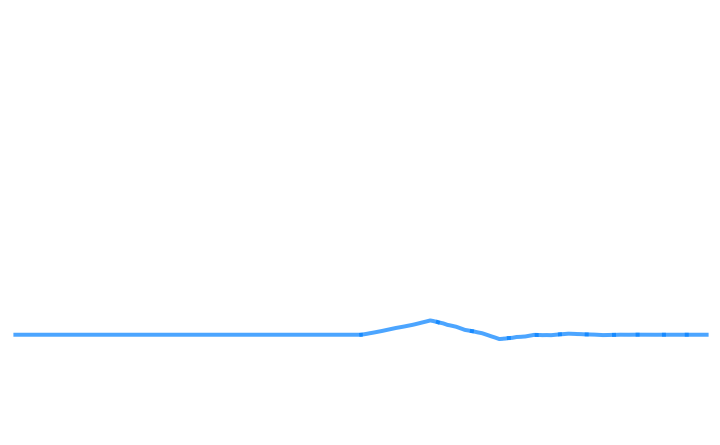}};

\node at (0,0) {\includegraphics[scale=0.3]{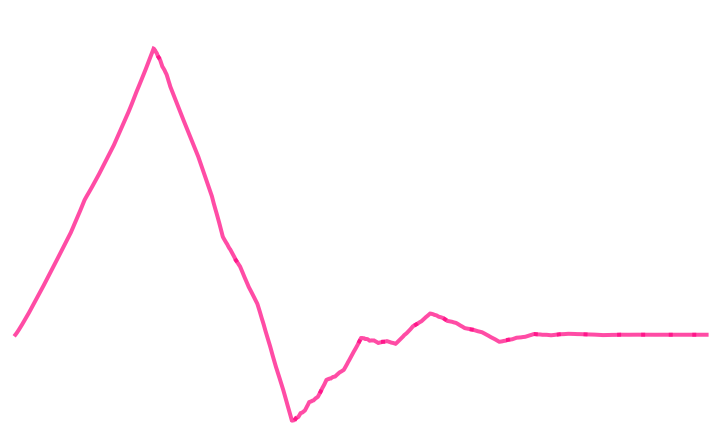}};
\node at (2.47,0) {$=$};
\node at (5,0) {\includegraphics[scale=0.3]{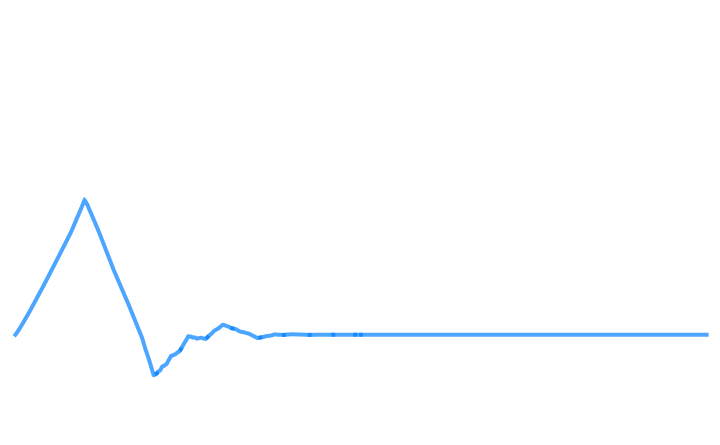}};
\node at (7.35,0) {$+$};
\node at (9.7,0) {\includegraphics[scale=0.3]{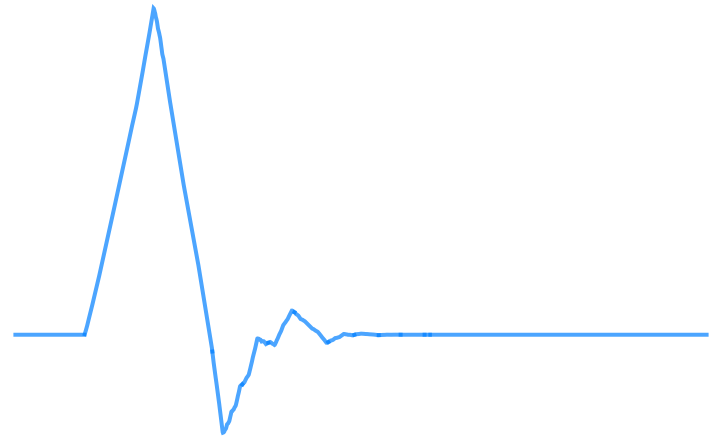}};
\node at (12.05,0) {$+$};
\node at (14.4,0) {\includegraphics[scale=0.3]{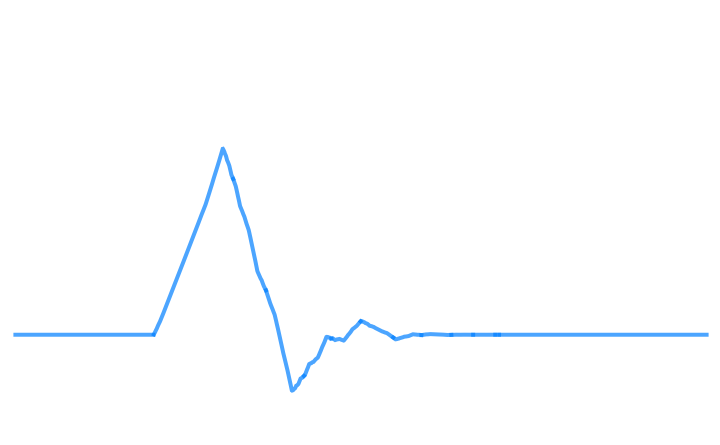}};

\node at (-.2,.2) {$s_0^r(x)$};
\node at (4.6,.1) {$h_0s^{r+1}_0(x)$};
\node at (9.8,.1) {$h_1s^{r+1}_1(x)$};
\node at (14.8,.1) {$h_2s^{r+1}_2(x)$};
\node at (4.6,-2.4) {$h_3s^{r+1}_3(x)$};
\node at (9.8,-2.4) {$h_4s^{r+1}_4(x)$};
\node at (14.8,-2.4) {$h_5s^{r+1}_5(x)$};

\foreach \cx/\cy in {0/0, 5/0, 9.7/0, 14.4/0, 5/-2.5, 9.7/-2.5, 14.4/-2.5} {
    \draw[->, gray, very thin] ({\cx-2},{\cy-0.6}) -- ({\cx+2.1},{\cy-0.6});
    \draw[->, gray, very thin] ({\cx-1.84},{\cy-0.75}) -- ({\cx-1.84},{\cy+1.1});
    \node[gray, font=\scriptsize] at ({\cx+2.27},{\cy-0.6}) {$x$};
    \foreach \i in {0,1,2,3,4,5} {
        \draw[gray, very thin] ({\cx-1.84+0.737*\i},{\cy-0.68}) -- ({\cx-1.84+0.737*\i},{\cy-0.52});
        \node[gray, font=\scriptsize] at ({\cx-1.84+0.737*\i},{\cy-0.92}) {\i};
    }
}

\end{tikzpicture}

\end{center}
\end{widetext}
which is quite non-trivial! In practice, we need to compute integrals of products of $s_n^r$, which can be obtained directly from the fixed-point equation. This is all we need to present our variational method at a fixed resolution.

\section{MPS at fixed resolution}\label{sec:wMPSfixed}
We now make our variational MPS manifold explicit, illustrate the minimization on the Lieb-Liniger model, and compute its energy density and correlation functions.

\subsection{The variational manifold}\label{sec:variational}
\begin{figure}
    \centering
    \begin{tikzpicture}

\begin{scope}[scale=1.6, xshift=-5, yshift=-3]
    \draw[
        thick,
        fill=violet!20,
        draw=violet!20
    ]
        (0.00,0.00) --
        (1.85,-0.90) --
        (3.58,-0.40) --
        (4.1,1.64) --
        (2.45,2.13) --
        (0.35,1.66) --
        cycle;

\end{scope}

\begin{scope}[scale=1.25]
    \draw[
        thick,
        fill=violet!35,
        draw=violet!35
    ]
        (0.00,0.00) --
        (1.85,-0.90) --
        (3.58,-0.40) --
        (4.1,1.64) --
        (2.45,2.13) --
        (0.35,1.66) --
        cycle;
\end{scope}

\begin{scope}[scale=.9, xshift=10, yshift=4]
    \draw[
        thick,
        fill=violet!50,
        draw=violet!50
    ]
        (0.00,0.00) --
        (1.85,-0.90) --
        (3.58,-0.40) --
        (4.1,1.64) --
        (2.45,2.13) --
        (0.35,1.66) --
        cycle;

\end{scope}

\begin{scope}[scale=.65, xshift=25, yshift=10]
    \draw[
        thick,
        fill=violet!65,
        draw=violet!65
    ]
        (0.00,0.00) --
        (1.85,-0.90) --
        (3.58,-0.40) --
        (4.1,1.64) --
        (2.45,2.13) --
        (0.35,1.66) --
        cycle;

\end{scope}

\begin{scope}[scale=.4, yscale=.65, xshift=80, yshift=30]
    \filldraw[
    fill=blue!27,
    draw=blue!27,
    thick
]
    (0,0)
    .. controls (1.5,-1.2) and (3,-0.8) ..
    (3.2,0.8)

    .. controls (3.4,2.2) and (2.2,3) ..
    (0.8,2.6)

    .. controls (-0.6,2.2) and (-1.2,1) ..
    (0,0);

\end{scope}

    \node at (5.6,2.2) {$\mathcal{H}$};
    \node at (4.6,1.8) {$\mathcal{H}^{r+1}$};
    \node at (3.6,1.4) {$\mathcal{H}^{r}$};
    \node at (2.75,1.05) {$\mathcal{H}^{r}_d$};
    \node at (1.65,.55) {$\mathcal{M}^{r,d,\chi}_{MPS}$};
    
\end{tikzpicture}
    \caption{The variational manifold. From the full Hilbert space $\mathcal{H}$, we first reduce our consideration to $\mathcal{H}^r$, the Fock space of resolution $r$ scaling modes at each site. We see that this space lies inside $\mathcal{H}^{r+1}$, and all higher resolution Hilbert spaces. Further reduction to $\mathcal{H}_d^r$ truncates each Fock space to $d-1$ excitations. Finally, we further reduce to the MPS submanifold $\mathcal{M}_\text{MPS}^{r,d,\chi}$.}
    \label{fig:var_mani}
\end{figure}
The Hilbert space $\mathcal{H}^r$ at fixed scale $r$ is defined as the tensor product of the Fock spaces associated to the modes
\begin{equation}\label{eq:a_def}
\hat a_n^r:= \int \upd x \, s_n^r(x) \, \hat\psi(x).\, 
\end{equation}
At this stage, each Fock space is still infinite dimensional. To use standard MPS algorithms we thus consider the truncated Fock space $\mathcal{H}^r_d\subset \mathcal{H}^r$ where each mode is populated with at most $d-1$ excitations. Since the particle \emph{density} is finite in the ground state, we expect that when $r$ is large, it is not necessary to increase $d$ beyond $3$ to get an arbitrarily good variational approximation. 

For example, for $I=[0, L]$ ($L$ integer) and periodic boundary conditions, one can fit $\ell = 2^r L$ such discrete truncated modes, and a basis of $\mathcal{H}_d^r$ is
\begin{equation}\label{eq:numstate}
  \ket{m_0,...,m_{\ell-1}}_r := \frac{(\hat a_0^{r\dag})^{m_0} (\hat a_1^{r\dag})^{m_1} \dots (\hat a_{\ell-1}^{r\dag})^{m_{\ell-1}}}{\sqrt{m_0! m_1! \dots m_{\ell-1}!}}\ket{0},
\end{equation}
where $m_0,\dots, m_{\ell-1}$ are each in $\{0,1,2,\dots,d-1\}$, and we have periodized the scaling functions.

Our final step is to consider an MPS manifold $\mathcal{M}_\text{MPS}^{r,d,\chi}$ at finite bond dimension $\chi$ within $\mathcal{H}_d^{r}$. This manifold is made of the states of the form:
\begin{equation}\label{eq:MPS_def}
\begin{aligned}
  \ket{A, r} &:= \!\!\sum_{m_0,\cdots, m_{\ell-1}=0}^{d-1} \!\!\!\!\!\!\!\!\!\!\text{Tr}(A^{m_0}_{0} ... A^{m_{\ell-1}}_{\ell-1}) \ket{m_0,... , m_{\ell-1}}_r \\
  &= \quad \vcenter{\hbox{\begin{tikzpicture}[line width=0.6pt]

    \node[rectangle, rounded corners,minimum width=0.5cm,minimum height=0.5cm, fill=blue!20, draw, line width=0.6pt] (A0) at (0,0) {$A_0$};

    \node[rectangle, rounded corners,minimum width=0.5cm,minimum height=0.5cm, fill=blue!20, draw, line width=0.6pt] (A1) at (1.3,0) {$A_1$};

    \node[rectangle, rounded corners,minimum width=0.5cm,minimum height=0.5cm, fill=blue!20, draw, line width=0.6pt] (A2) at (2.6,0) {$A_2$};

    \node at (3.75,0) {$\cdots$};

    \node[rectangle, rounded corners,minimum width=0.5cm,minimum height=0.5cm, fill=blue!20, draw, line width=0.6pt] (A3) at (4.9,0) {$A_{\ell-1}$};

    \draw (A0.east) -- (A1.west);
    \draw (A1.east) -- (A2.west);
    \draw (A2.east) -- (3.45,0);
    \draw (4.05,0) -- (A3.west);

    \draw (A0.south) -- (0,-0.7);
    \draw (A1.south) -- (1.3,-0.7);
    \draw (A2.south) -- (2.6,-0.7);
    \draw (A3.south) -- (4.9,-0.7);
    \node at (0,-1) {$m_0$};
    \node at (1.3,-1) {$m_1$};
    \node at (2.6,-1) {$m_2$};
    \node at (4.9,-1) {$m_{\ell-1}$};

    \node at (0.65,0.2) {$\chi$};

    \draw[rounded corners=6pt] (A3.east) -- (5.7,0) -- (5.7,0.85) -- (-0.8,0.85) -- (-0.8,0) -- (A0.west);

\end{tikzpicture}}}\;,
\end{aligned}
\end{equation}
where $A_k^{m}$ are $\ell\times d$ matrices of size $\chi\times \chi$ and we have used the standard tensor network notation~\cite{Bridgeman_Chubb_2017} on the second line.

In the translation invariant case, one can drop the $k$ dependency, and there are just $d$ matrices. They contain the variational parameters that one optimizes over to find the ground state. Further taking the thermodynamic limit, we obtain infinite MPSs
\begin{equation}\label{eq:iMPS_def}
  \ket{A,r} = \quad \vcenter{\hbox{\begin{tikzpicture}[line width=0.6pt]

    \node[rectangle, rounded corners,minimum width=0.5cm,minimum height=0.5cm, fill=blue!20, draw] (A1) at (0,0) {$A$};

    \node[rectangle, rounded corners,minimum width=0.5cm,minimum height=0.5cm, fill=blue!20, draw] (A2) at (1.1,0) {$A$};

    \node[rectangle, rounded corners,minimum width=0.5cm,minimum height=0.5cm, fill=blue!20, draw] (A3) at (2.2,0) {$A$};

    \node[rectangle, rounded corners,minimum width=0.5cm,minimum height=0.5cm, fill=blue!20, draw] (A4) at (3.3,0) {$A$};

    \node at (-1,0) {...};
    \node at (4.3,0) {...};

    \draw (-.8,0) -- (A1.west);
    \draw (A1.east) -- (A2.west);
    \draw (A2.east) -- (A3.west);
    \draw (A3.east) -- (A4.west);
    \draw (A4.east) -- (4.1,0);
    \draw (A1.south) -- (0,-0.7);
    \draw (A2.south) -- (1.1,-0.7);
    \draw (A3.south) -- (2.2,-0.7);
    \draw (A4.south) -- (3.3,-0.7);

    \node at (1.65,0.2) {$\chi$};

    \node at (1.1,-1) {{$m_k$}};

\end{tikzpicture}
}} \, ,
\end{equation}
which are a submanifold of the original Hilbert space $\mathcal{H}$ as illustrated in Fig. \ref{fig:var_mani}. We call these states $\ket{A,r}$ wavelet-MPS or wMPS\footnote{Here, we use ``wavelet'' in the generic sense it has, \textit{e.g.}, in ``wavelet theory''. At this stage at least, wMPS do not use wavelet \emph{functions} in their construction (only the scaling functions).}.

\subsection{Evaluating the Hamiltonian}
We now aim to evaluate the expectation value of operators, and in particular the Hamiltonian (density) on the states of our manifold. To this end, we first note that for any normal-ordered polynomial $P$ in $\hat{\psi}^\dagger$ and $\hat{\psi}$, and a state $\ket{\xi} \in \mathcal{H}^r$, we have
\begin{equation}\label{eq:normal_ordered_at_r}
  \bra{\xi} :\! P(\hat{\psi}^\dagger,\hat{\psi})\!: \ket{\xi} = \bra{\xi} :\!P(\hat{\psi}^\dagger_r,\hat{\psi}_r)\!: \ket{\xi}, 
\end{equation}
with $\hat{\psi}_r$ the continuous mode operator projected at the resolution $r$:
\begin{equation}
  \hat{\psi}_r(x) := \sum_{n\in\mathbb{Z}} s_n^r(x)\,  \hat a_n^r \, .
  \label{eq:test}
\end{equation}
This is intuitive, and follows formally from the definition of the modes at scale $r$ \eqref{eq:a_def} and the canonical commutation relation $[\hat{\psi}(x),\hat{\psi}^\dagger(y)]=\delta(x-y)$.

This implies that we have $\bra{\xi} H \ket{\xi} = \bra{\xi} H^r \ket{\xi}$ for  $\ket{\xi} \in \mathcal{H}^r$ with the Hamiltonian at resolution $r$:
\begin{equation}
  \begin{split}
    H^r &=\int_I \upd x \sum_{n,m}\hat a_n^{r\dag} \hat a_m^r \partial_x s_n^r(x) \partial_x s_m^r(x)  \\
    &+ c \int_I \upd x\sum_{n,m,l,k}\hat a_n^{r\dag} \hat a_m^{r\dag} \hat a_l^r \hat a_k^r s_n^r(x) s_m^r(x) s_l^r(x) s_k^r(x) \\
    &- \mu \int_I \upd x\sum_{n,m}\hat a_n^{r\dag} \hat a_m^r s_n^r(x) s_m^r(x) \label{eq:fullH}\,.
  \end{split}
\end{equation}
Inverting the sum and the integral over $x$ gives
\begin{equation}
  \begin{split}
    H^r&= 2^{2r}\sum_{n,m} K_{m-n} \hat a_n^{r\dag} \hat a_m^r  \\
       &+ 2^r c \sum_{n,m,l,k} \Gamma^4_{m-n, l-n, k-n} \hat a_n^{r\dag} \hat a_m^{r\dag} \hat a_l^r \hat a_k^r  \\
    &- \mu \sum_n \hat a_n^{r\dag} \hat a_n^r,
  \end{split}
\end{equation}
where the kinetic coefficients $K$ are given by the integrals over the derivative of scaling functions
\begin{align}\label{eq:K}
  \int \upd x\, \partial_x s_n^r(x) \partial_x s_m^r(x) 
    &= 2^{2r} \int \upd x\, \partial_x s(x) \partial_x s_{m-n}(x) \nonumber \\
    &:= 2^{2r} K_{m-n},
\end{align}
and the four-point interaction coefficients $\Gamma^4$ are defined from the integral of $4$ scaling functions
\begin{equation}\label{eq:Gamma4}
  \begin{split}
    &\int \upd x \; s_n^r(x) s_m^r(x) s_l^r(x) s_k^r(x)   \\ &\quad =2^{r} \int \upd x\; s(x) s_{m-n}(x) s_{l-n}(x) s_{k-n}(x)\\ 
    &\quad   := 2^r \Gamma^4_{m-n, l-n, k-n}. 
    \end{split}
\end{equation}
These coefficients can be evaluated using the fixed-point equation~\eqref{eq:RGs} \cite{PhysRevD.101.096004}, as detailed in the Appendix. Because $s$ is compactly supported, both the kinetic and interaction coefficients are zero for site separations larger than $N-1$ (for the interaction term this means the separation between any pair of sites must be less than $N-1$ for the term to be non-zero). This implies that $H^r$ is local, albeit with interactions beyond nearest neighbor.

In what follows, we consider Daubechies wavelets of order $N=6,8$ and compute the coefficients $K$ and $\Gamma^4$ for each order once and for all. Since it is local, we can then express $H^r$ as a Matrix Product Operator (MPO), or an infinite MPO \cite{PhysRevB.102.035147} in the thermodynamic limit:
\begin{equation}\label{eq:Hr_mpo}
  H^r = \quad \vcenter{\hbox{\begin{tikzpicture}[line width=0.6pt]

    \node[rectangle, rounded corners,minimum width=0.5cm,minimum height=0.5cm, fill=orange!20, draw] (A1) at (0,0) {$h$};

    \node[rectangle, rounded corners,minimum width=0.5cm,minimum height=0.5cm, fill=orange!20, draw] (A2) at (1.1,0) {$h$};

    \node[rectangle, rounded corners,minimum width=0.5cm,minimum height=0.5cm, fill=orange!20, draw] (A3) at (2.2,0) {$h$};

    \node[rectangle, rounded corners,minimum width=0.5cm,minimum height=0.5cm, fill=orange!20, draw] (A4) at (3.3,0) {$h$};

    \node at (-1,0) {...};
    \node at (4.3,0) {...};

    \draw (-.8,0) -- (A1.west);
    \draw (A1.east) -- (A2.west);
    \draw (A2.east) -- (A3.west);
    \draw (A3.east) -- (A4.west);
    \draw (A4.east) -- (4.1,0);

    \node at (1.65,0.45) {\small $\chi_\text{MPO}$};

    \draw (A1.north) -- (0,0.7);
    \draw (A2.north) -- (1.1,0.7);
    \draw (A3.north) -- (2.2,0.7);
    \draw (A4.north) -- (3.3,0.7);

    \draw (A1.south) -- (0,-0.7);
    \draw (A2.south) -- (1.1,-0.7);
    \draw (A3.south) -- (2.2,-0.7);
    \draw (A4.south) -- (3.3,-0.7);

\end{tikzpicture}}}\,.
\end{equation}
This MPO encoding is exact, which is crucial to keep our method variational. However, because the interactions are not nearest neighbor, especially in the 4-point interaction term, the MPO bond dimension is quite large at $\chi_\text{MPO}=702$ for $N=6$ and $\chi_\text{MPO}=3306$ for $N=8$, and would grow even larger for larger $N$. Despite this large MPO bond dimension, calculations remain numerically tractable.

\subsection{Energy density}\label{sec:error}
\begin{figure}
    \centering
    \includegraphics[width=\linewidth]{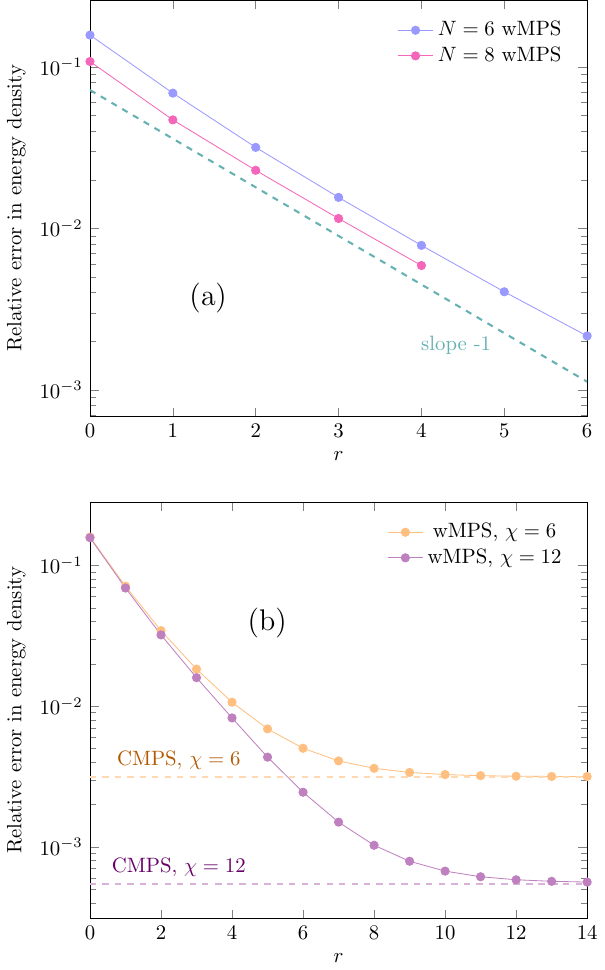}
    \caption{Relative error in the wMPS ground state energy density as a function of the resolution $r$, for the Lieb-Liniger model with $\mu=1,c=8$ and an MPS with $d=3$ states in the Fock space on each physical leg. (a) We compare $N=6$ and $N=8$ wavelet MPS with bond dimension $\chi=16$, which both scale linearly with the lattice spacing. (b) We compare results for wMPS with bond dimension $\chi=6,12$, and larger $r$ for $N=6$, to witness the saturation as a function of $r$ to the CMPS values. For $r\geq 5$, the optimization is done with the iterative refinement method of sec. \ref{sec:refinement}.}
    \label{fig:error}
\end{figure}

We are now ready to minimize the energy density of the Lieb-Liniger model over our variational manifold at fixed resolution $r$. This step is now completely standard, and we can use any off-the-shelf algorithm. In practice, we use \texttt{MPSKit.jl} \cite{Devos_MPSKit_2026} to run the variational uniform matrix product states algorithm (VUMPS) \cite{PhysRevB.97.045145, 10.21468/SciPostPhysLectNotes.7}, before switching to gradient descent on the Grassmann manifold \cite{10.21468/SciPostPhys.10.2.040} as the state nears convergence. This switch method is standard, and typically motivated by the fact that it is faster for certain benchmark models \cite{10.21468/SciPostPhys.10.2.040}. In our case, the primary motivation is that VUMPS, though fast, becomes unstable at high resolutions when nearing convergence. For $r\geq 6$, energy minimization becomes difficult no matter the method, which motivates the introduction of iterative refinement in Sec. \ref{sec:refinement}.

We compare the relative error in the ground state energy density we obtain with the exact Bethe-Ansatz solution \cite{10.21468/SciPostPhys.3.1.003, PhysRevB.100.081110} in Fig.~\ref{fig:error}. We first consider a bond dimension $\chi=16$, which is large enough that the error for $r\leq 6$ is dominated by the finite lengthscale introduced. We observe that the relative error then decays exponentially in $r$, \textit{i.e.} linearly in the typical lengthscale $2^{-r}$. Although this is the same decay one could find with more naive discretization methods, our method has the advantage of being truly variational. One may suspect that one could get a faster decay with smoother scaling functions. The $D6$ scaling functions have Hölder exponent $\alpha=1.0878$~\cite{Daubechies_1992}, so they are just above the threshold of differentiability. The $D8$ scaling functions have $\alpha=1.6179$, and are thus substantially more regular (although not twice differentiable). As we can see in Fig. \ref{fig:error}, the convergence rate seems to be unrelated to this regularity.

It might be worth trying even more regular functions, but we suspect that it is not possible to get faster convergence without abandoning our approach consisting in first defining an orthonormal family of $L^2(\mathbb{R})$, and then building local Fock spaces from it. For example, in \cite{PhysRevLett.128.230401}, the authors obtain a better error scaling by defining the $n$-body wavefunctions on small intervals separately (namely, states with two excitations are not made from tensor products of the one-excitation states present in the basis already). In that case, the Hilbert space they consider, analogous to our $\mathcal{H}^r_d$, is not a truncated Fock space. This requires a subspace construction tailored to the Hamiltonian considered (and in principle, rebuilt for every coupling). Further, in the case of \cite{PhysRevLett.128.230401}, it makes the method not variational since one needs to add a formally infinite penalty term to connect intervals.

Our solution to the slow convergence problem is different. Since the error decreases exponentially in $r$, we can reach any precision desired provided we can increase $r$ arbitrarily, through iterative refinements, without re-optimizing from a random state every time. This is the method we present in Sec. \ref{sec:refinement}, and that leverages the full power of the multi-resolution analysis and makes use of wavelets.

We now compare our results with CMPS values obtained with \texttt{CMPSKit.jl}~\cite{tuybens2022}. To this end, we fix $\chi$ and push to much larger $r$ (using the refinement method described in the next section). As expected the relative error in the energy density reaches a plateau, where finite entanglement is the bottleneck. Interestingly, this plateau is precisely the CMPS value. This leads us to conjecture that, in fact, our wMPS at $r\rightarrow +\infty$ are \emph{exactly} CMPS.

\subsection{Correlation functions}\label{sec:correlators}
\begin{figure}
    \centering
    \includegraphics[width=.48\textwidth]{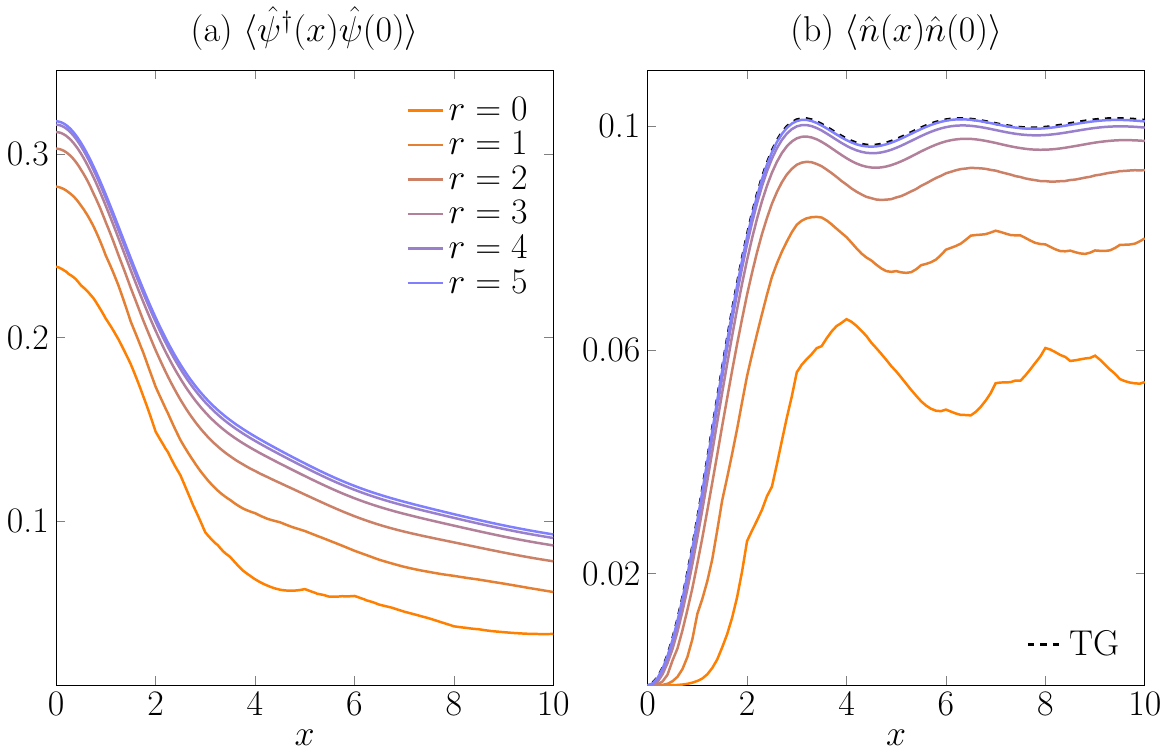}
    \caption{Observables in the wMPS ground state with $\mu=1,\chi=16,d=3$, at different resolutions, near the Tonks-Girardeau limit at $c=200$.}
    \label{fig:correlators}
\end{figure}

Once the approximate ground state $\ket{A,r}$ has been obtained, computing the expectation value of observables is straightforward. We insist that, just like with the Hamiltonian, the observables are \emph{not} discretized or approximated in any way. In particular, unlike with a naive lattice discretization, we can compute correlation functions at arbitrary real separation $x$, even if the scale $r$ is fixed. To this end, we use eq.~\eqref{eq:normal_ordered_at_r} to express the exact expectation value of $\ket{A,r}$ as an expectation value involving only the scaling modes $\hat a_n^r$. For example, for the two-point function, we have
\begin{widetext}
\begin{equation}
  \begin{split}
    \bra{A, r}\hat\psi^\dag(x)\hat\psi(0)\ket{A,r} &= \bra{A, r}\hat\psi^\dag_r(x)\hat\psi_r(0)\ket{A,r} \\
                                                   &=2^r\sum_{n=\lceil 2^rx \rceil -N+1}^{\lfloor 2^r x \rfloor} \sum_{m=-N+2}^{-1} s(2^rx-n)  s(-m) \bra{A,r}\hat a_n^{r\dag}\hat a_m^r\ket{A,r}.
  \end{split}
\end{equation}
\end{widetext}
Crucially, this is a finite sum, because $s$ has compact support. The density-density correlator $\bra{A, r}\hat\psi^\dag(x)\hat\psi^\dag(0)\hat\psi(x)\hat\psi(0)\ket{A, r}$ can be computed in an analogous way.

The two-point and density-density correlation functions are  shown for different resolutions in Fig. \ref{fig:correlators}. We deliberately chose small $r$ here so that the finite $r$ artifacts are visible. We get the expected behavior at sufficiently large $r$. In particular, for the density-density correlator, our results become visually indistinguishable from the exact Tonks-Girardeau (TG) limit $c\rightarrow +\infty$, with the Friedel oscillations visible already at smaller $r$. 

Interestingly, for the smallest $r$, we see that the correlators are not just coarser versions of the higher $r$ ones (with the rough pattern of the $D6$ scaling function imprinted on top), but are actually globally shifted. This is expected of a true variational method: we are not just coarse graining the exact solution, we are minimizing over coarser states, and thus reaching slightly different approximate ground states for each $r$.

\section{Multi-resolution analysis}\label{sec:refinement}
So far we have only considered wMPS at fixed resolution: each time the resolution is increased, we start from a random initial state, and run the full ground-state optimization routine. However, by construction $\mathcal{H}^{r}\subset \mathcal{H}^{r+1}$, and thus, at least for some $\tilde{\chi}\geq \chi$, we have $\mathcal{M}_\text{MPS}^{r,d,\chi}\subset \mathcal{M}_\text{MPS}^{r+1,d,\tilde{\chi}}$. Thus, in analogy with bond-space expansion techniques \cite{White2005,Hubig2015,10.21468/SciPostPhysLectNotes.7,Gleis2023, Li2024}, we should be able to start from an approximate ground state at resolution $r$, and promote it to an approximate ground state at resolution $r+1$ with at most the same energy density. This refined approximate ground state could then be taken as a starting point for energy minimization, which would converge far quicker than when starting from a random state.

This refinement can be done using the full strength of multi-resolution analysis, which is what we do in this section. We first introduce the second half of the theory of Daubechies wavelets that we had postponed, then present our strategy to implement the refinement, explain in detail the different steps, and finally test it. Quite unexpectedly, we find that refining without optimizing further already substantially lowers the energy density.

\subsection{Wavelet functions}

Using the refinement relation \eqref{eq:WTs}, one can write modes at a coarse scale in terms of modes at a finer scale, and thus naturally write a state in $\mathcal{H}^r$ as a state in $\mathcal{H}^{r+1}$. We will later need to implement this embedding locally, with a unitary circuit. When doing so, it is helpful to understand the orthogonal complement of $\mathcal{H}^r$ within $\mathcal{H}^{r+1}$, and this is when wavelet functions become helpful.

The wavelet space $\mathcal{W}_r$ is simply defined as the orthogonal complement of $\mathcal{S}_r$: $\mathcal{S}_{r+1} = \mathcal{S}_r \oplus \mathcal{W}_r$. Applying this recursively, one covers $L^2(\mathbb{R})$
\begin{equation}
  L^2(\mathbb{R}) = \mathcal{S}_r\oplus \mathcal{W}_r \oplus \mathcal{W}_{r+1}\oplus \mathcal{W}_{r+2}\oplus \dots \; .
\end{equation}
The subspace $\mathcal{W}_r$ is spanned by a new set of functions, $\{w_n^r(x)\} |_{n\in \mathbb{Z}}$, the wavelet functions. They are orthogonal to the scaling functions, satisfy $w_n^r(x) = 2^{r/2}w(2^r x - n)$, and can be written explicitly in terms of the scaling functions:
\begin{equation}\label{eq:WTw}
    w_n^r(x) = \sum_{i=0}^{N-1}g_{i} s_{2n+i}^{r+1}(x),
\end{equation}
where $g_i := (-1)^i h_{N-1-i}$ are the wavelet function weights. 

At resolution $r+1$, the orthonormal families $\{s_n^{r+1}(x)\} |_{n \in \mathbb{Z}}$ and $\{s_n^r(x)\} |_{n \in \mathbb{Z}} \bigcup \{w_n^r(x)\} |_{n \in \mathbb{Z}}$ span the same subspace. The explicit relation is given by the inverse wavelet transform: 
\begin{equation}\label{eq:IWT}
\begin{split}
    s_{2n}^{r+1}(x) =& \sum_{i=0}^{\frac{N}{2}-1} h_{2i} s_{n-i}^r(x) + g_{2i} w_{n-i}^r(x) \\ 
    s_{2n+1}^{r+1}(x) =& \sum_{i=0}^{\frac{N}{2}-1} h_{2i+1} s_{n-i}^r(x) + g_{2i+1} w_{n-i}^r(x).
\end{split}
\end{equation}
The requirement that scaling functions can locally represent polynomials implies that the moments of the wavelet functions up to order $N/2-1$ are zero:
\begin{equation}
  \int_\mathbb{R}\upd x\, x^k w(x) = 0,\text{ for }0 \leq k \leq N/2-1.
\end{equation}
This can be easily seen from eq.~\eqref{eq:polys} and the fact that $\int \upd x s_n(x)w(x) = 0$ for all $n \in \mathbb{Z}$. Just like the scaling function $s$, the wavelet function $w$ is compactly supported on $[0,N-1]$, becomes more regular as $N$ increases, and has continuous first derivative for $N \geq 6$.

Moving from $L^2(\mathbb{R})$ onto Fock spaces, one can introduce the wavelet modes 
\begin{equation}
  \hat b_n^r= \int_\mathbb{R}\upd x\, w_n^r(x) \hat\psi(x) \, ,
\end{equation}
which commute with $\hat a_n^r$ and verify
\begin{equation}
     [\hat b^r_n, \hat b^{q\dag}_m] = \delta_{rq}\delta_{nm},\,\,\,\,\,\, [\hat b^r_n, \hat b^q_m] = [\hat b^{r\dag}_n, \hat b^{q\dag}_m] = 0.
 \end{equation}
As was done in~\cite{PhysRevA.92.032315}, one can now express the continuous annihilation operator $\hat\psi(x)$ exactly in terms of scaling and wavelet modes: 
\begin{equation}\label{eq:DWT}
     \hat\psi(x) = \sum_n \left(\hat a_n^r s_n^r(x) + \sum_{j=0}^{\infty} \hat b_n^{r+j} w_n^{r+j}(x) \right)\, .
 \end{equation}

\subsection{Strategy}
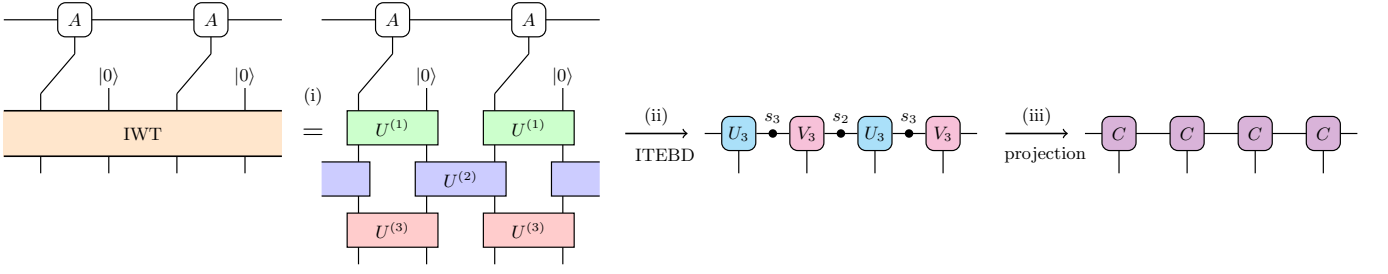
\begin{figure*}
    \centering
    \begin{tikzpicture}[line width=0.6pt,
    Abox/.style={rectangle, rounded corners, minimum width=0.6cm, minimum height=0.6cm, draw, fill=white},
    U1box/.style={rectangle, minimum width=1.6cm, minimum height=0.6cm, draw, fill=green!20},
    U2box/.style={rectangle, minimum width=1.6cm, minimum height=0.6cm, draw, fill=blue!20},
    U3box/.style={rectangle, minimum width=1.6cm, minimum height=0.6cm, draw, fill=red!20},
    Cbox/.style={rectangle, rounded corners, minimum width=0.6cm, minimum height=0.6cm, draw, fill=violet!30},
    dot/.style={circle, fill, inner sep=1.5pt},
]

\begin{scope}
    \draw (-0.45,1.6) -- (4.45,1.6);
    \foreach \x in {0.8,3.2}{ \node[Abox] at (\x,1.6) {$A$}; }
    \foreach \xc/\xs/\xv in {0.8/0.2/1.4, 3.2/2.6/3.8}{
        \draw (\xc,1.3) -- (\xc,1.0);
        \draw (\xc,1.0) -- (\xs,0.3);
        \node at (\xv,0.6) {$\ket{0}$};
    }
    \foreach \x in {0.2,2.6}{ \draw (\x,0) -- (\x,0.3); }
    \foreach \x in {1.4,3.8}{ \draw (\x,0) -- (\x,0.4); }
    \node[rectangle, minimum width=4.9cm, minimum height=0.8cm, fill=orange!20] at (2.0,-0.4) {IWT};
    \draw[thick] (-0.45,0) -- (4.45,0);
    \draw[thick] (-0.45,-0.8) -- (4.45,-0.8);
    \foreach \x in {0.2,1.4,2.6,3.8}{ \draw (\x,-0.8) -- (\x,-1.1); }
\end{scope}

\node at (5.0,-0.4) {\Large $=$};
\node at (5.0,0.25) {(i)};

\begin{scope}[xshift=5.6cm]
    \draw (-0.45,1.6) -- (4.45,1.6);
    \foreach \x in {0.8,3.2}{ \node[Abox] at (\x,1.6) {$A$}; }
    \foreach \xc/\xs/\xv in {0.8/0.2/1.4, 3.2/2.6/3.8}{
        \draw (\xc,1.3) -- (\xc,1.0);
        \draw (\xc,1.0) -- (\xs,0.3);
        \node at (\xv,0.6) {$\ket{0}$};
    }
    \foreach \x in {0.2,2.6}{ \draw (\x,0) -- (\x,0.3); }
    \foreach \x in {1.4,3.8}{ \draw (\x,0) -- (\x,0.4); }
    \foreach \x in {0.8,3.2}{ \node[U1box] at (\x,-0.3) {$U^{(1)}$}; }
    \foreach \x in {0.2,1.4,2.6,3.8}{ \draw (\x,-0.6) -- (\x,-0.9); }
    \node[U2box] at (2.0,-1.2) {$U^{(2)}$};
    \draw[fill=blue!20] (-0.45,-0.9) -- (0.4,-0.9) -- (0.4,-1.5) -- (-0.45,-1.5);
    \draw[fill=blue!20] (4.45,-0.9) -- (3.6,-0.9) -- (3.6,-1.5) -- (4.45,-1.5);
    \foreach \x in {0.2,1.4,2.6,3.8}{ \draw (\x,-1.5) -- (\x,-1.8); }
    \foreach \x in {0.8,3.2}{ \node[U3box] at (\x,-2.1) {$U^{(3)}$}; }
    \foreach \x in {0.2,1.4,2.6,3.8}{ \draw (\x,-2.4) -- (\x,-2.7); }
\end{scope}

\draw[->, thick] (10.5,-0.4) -- (11.6,-0.4);
\node at (11.05,-0.1) {(ii)};
\node[font=\small] at (11.2,-0.78) {ITEBD};

\begin{scope}[shift={(12.3,-0.4)}]
    \node[Abox, fill=cyan!30]    (e1) at (0.2,0) {$U_3$};
    \node[Abox, fill=magenta!30] (e2) at (1.4,0) {$V_3$};
    \node[Abox, fill=cyan!30]    (e3) at (2.6,0) {$U_3$};
    \node[Abox, fill=magenta!30] (e4) at (3.8,0) {$V_3$};
    \draw (-0.4,0) -- (e1.west); \draw (e1.east) -- (e2.west); \draw (e2.east) -- (e3.west);
    \draw (e3.east) -- (e4.west); \draw (e4.east) -- (4.4,0);
    \node[dot] at (0.8,0) {}; \node at (0.8,0.27) {$s_3$};
    \node[dot] at (2.0,0) {}; \node at (2.0,0.27) {$s_2$};
    \node[dot] at (3.2,0) {}; \node at (3.2,0.27) {$s_3$};
    \foreach \x in {0.2,1.4,2.6,3.8}{ \draw (\x,-0.3) -- (\x,-0.7); }
\end{scope}

\draw[->, thick] (17.2,-0.4) -- (18.3,-0.4);
\node at (17.75,-0.1) {(iii)};
\node[font=\small] at (17.9,-0.78) {projection};

\begin{scope}[shift={(19.0,-0.4)}]
    \draw (-0.4,0) -- (4.4,0);
    \foreach \x in {0.2,1.4,2.6,3.8}{ \node[Cbox] at (\x,0) {$C$}; }
    \foreach \x in {0.2,1.4,2.6,3.8}{ \draw (\x,-0.3) -- (\x,-0.7); }
\end{scope}

\end{tikzpicture}
    \caption{Illustration of the three steps of the refinement algorithm: (i) writing the inverse wavelet transform as a brick-wall circuit, (ii) contracting it with ITEBD, (iii) projecting onto translation invariant MPSs.}
    \label{fig:refalg}
\end{figure*}
We now give the general strategy of our refinement method, to motivate the subsequent technical subsections. For simplicity, we restrict to $N=6$ Daubechies wavelets from now on.

We start from a matrix product state $\ket{A, r}$ optimized at resolution $r$. Our refinement algorithm is presented in Fig. \ref{fig:refalg} and involves three steps:
\begin{enumerate} 
    \item[i.] Applying the IWT \eqref{eq:IWT}, written as a quantum circuit, to $\ket{A, r}$ with the wavelet modes in the vacuum,
    \item[ii.] Contracting the resulting tensor network with $3$ steps of the infinite time-evolving block decimation (ITEBD) algorithm \cite{PhysRevLett.98.070201},     
    \item[iii.] Restoring translation invariance, to get back a wMPS state $\ket{C, r+1}$.
\end{enumerate}
After step (ii), the MPS has a two-site unit cell, but should be the same state as the original one, up to numerical truncation errors. However, the third step does change the state, and is not a pure change of gauge. Intuitively, since the ground state is translation invariant, forcing translation invariance at a finer scale could be helpful, and this is indeed what we observe.

\subsection{The inverse wavelet transform as a quantum circuit}

Writing the inverse wavelet transform as a quantum circuit is quite a standard step, that can be found in various forms in the literature (see \textit{e.g.}~\cite{PhysRevA.97.052314}). We present it in detail here both for completeness, and also because we need a more explicit form. Namely, we need not only a formal expression at the mode level (which suffices for free theories), but the coefficients in Fock space.

The IWT is given in eq.~\eqref{eq:IWT} in terms of scaling and wavelet functions. It implies the following Bogoliubov transform at the level of modes
\begin{eqnarray}\label{eq:IWTop}
    \hat a_{2n}^{r+1} =& h_0 \hat a_{n}^r& + h_2 \hat a_{n-1}^r + h_4 \hat a_{n-2}^r \nonumber \\
    &&+ g_0 \hat b_{n}^r + g_2 \hat b_{n-1}^r + g_4 \hat b_{n-2}^r \nonumber \\ 
    \hat a_{2n+1}^{r+1} =& h_1 \hat a_{n}^r& + h_3 \hat a_{n-1}^r + h_5 \hat a_{n-2}^r \nonumber \\
    &&+ g_1 \hat b_{n}^r + g_3 \hat b_{n-1}^r + g_5 \hat b_{n-2}^r\, ,
\end{eqnarray}
or in abstract form
\begin{equation}
  \left(\begin{array}{c}
      \cdots \\
      \hat{a}_{2n}^{r+1}\\ 
      \hat{a}_{2n+1}^{r+1}\\  
    \cdots 
\end{array}\right) = \mathcal{U}
\left(\begin{array}{c}
    \cdots\\
    \hat{a}_{n}^{r}\\
    \hat{b}_{n}^{r}\\
    \cdots 
  \end{array}\right)\,.
\end{equation}
The first step is to write this unitary transformation $\mathcal{U}$ in mode space as a local brick-wall circuit. For $N=6$, it has depth $3$:
\begin{equation}\label{eq:bogo_as_brickwall}
  \mathcal{U}= \vcenter{\hbox{\begin{tikzpicture}[line width=0.6pt]

\node[rectangle,minimum width=6.7cm,minimum height=.9cm, fill=cyan!20] at (3.2,-0.3) {};
\node[rectangle,minimum width=4.3cm,minimum height=1.1cm, fill=cyan!20] at (3.2,-1.1) {};
\node[rectangle,minimum width=1.9cm,minimum height=1.1cm, fill=cyan!20] at (3.2,-2) {};

\draw (0.2,.6) node {${\hat a_{n-2}^r}$};
\draw (1.4,.6) node {$\hat b_{n-2}^r$};

\draw (2.6,.6) node {${\hat a_{n-1}^r}$};
\draw (3.8,.6) node {$\hat b_{n-1}^r$};

\draw (5,.6) node {${\hat a_{n}^r}$};
\draw (6.2,.6) node {$\hat b_{n}^r$};

\node[rectangle,minimum width=1.6cm,minimum height=0.6cm, draw,fill=green!20] (a11) at (0.8,-0.3) {$u^{(1)}$};
\node[rectangle,minimum width=1.6cm,minimum height=0.6cm, draw,fill=green!20] (a12) at (3.2,-0.3) {$u^{(1)}$};
\node[rectangle,minimum width=1.6cm,minimum height=0.6cm, draw,fill=green!20] (a13) at (5.6,-0.3) {$u^{(1)}$};

\draw (0.2, 0) -- (0.2, 0.3);
\draw (1.4, 0) -- (1.4, 0.3);
\draw (2.6, 0) -- (2.6, 0.3);
\draw (3.8, 0) -- (3.8, 0.3);
\draw (5.0, 0) -- (5.0, 0.3);
\draw (6.2, 0) -- (6.2, 0.3);

\draw (0.2, -0.6) -- (0.2, -0.9);
\draw (1.4, -0.6) -- (1.4, -0.9);
\draw (2.6, -0.6) -- (2.6, -0.9);
\draw (3.8, -0.6) -- (3.8, -0.9);
\draw (5.0, -0.6) -- (5.0, -0.9);
\draw (6.2, -0.6) -- (6.2, -0.9);

\node[rectangle,minimum width=1.6cm,minimum height=0.6cm, draw,fill=blue!20] (a21) at (2.0,-1.2) {$u^{(2)}$};
\node[rectangle,minimum width=1.6cm,minimum height=0.6cm, draw,fill=blue!20] (a22) at (4.4,-1.2) {$u^{(2)}$};
\draw[fill=blue!20] (0, -0.9) -- (0.4, -0.9) -- (0.4,-1.5) -- (0,-1.5);
\draw[fill=blue!20] (6.4, -0.9) -- (6, -0.9) -- (6,-1.5) -- (6.4,-1.5);

\draw (0.2, -1.5) -- (0.2, -1.8);
\draw (1.4, -1.5) -- (1.4, -1.8);
\draw (2.6, -1.5) -- (2.6, -1.8);
\draw (3.8, -1.5) -- (3.8, -1.8);
\draw (5.0, -1.5) -- (5.0, -1.8);
\draw (6.2, -1.5) -- (6.2, -1.8);

\node[rectangle,minimum width=1.6cm,minimum height=0.6cm, draw,fill=red!20] (a31) at (0.8,-2.1) {$u^{(3)}$};
\node[rectangle,minimum width=1.6cm,minimum height=0.6cm, draw,fill=red!20] (a32) at (3.2,-2.1) {$u^{(3)}$};
\node[rectangle,minimum width=1.6cm,minimum height=0.6cm, draw,fill=red!20] (a33) at (5.6,-2.1) {$u^{(3)}$};

\draw (0.2, -2.4) -- (0.2, -2.7);
\draw (1.4, -2.4) -- (1.4, -2.7);
\draw (2.6, -2.4) -- (2.6, -2.7);
\draw (3.8, -2.4) -- (3.8, -2.7);
\draw (5.0, -2.4) -- (5.0, -2.7);
\draw (6.2, -2.4) -- (6.2, -2.7);

\draw (2.6,-3.05) node {${\hat a_{2n}^{r+1}}$};
\draw (3.8,-3.05) node {$\hat a_{2n+1}^{r+1}$};

\end{tikzpicture}}}\, .
  \end{equation}
The three unitary two-by-two matrices---$u^{(i)}$ for $i=1,2,3$---acting in operator space, can be determined from eq.~\eqref{eq:IWTop} and one gets:
\begin{equation}\label{eq:ai}
\begin{split}
    u^{(1)} =& \frac{1}{\sqrt{h_1^2+h_4^2}}\begin{pmatrix}
        h_1 & -h_4 \\
        h_4 & h_1
    \end{pmatrix}, \\
    u^{(2)} =& -\begin{pmatrix}
        \sqrt{h_2^2+h_3^2} & -\sqrt{1-h_2^2-h_3^2} \\
        \sqrt{1-h_2^2-h_3^2} & \sqrt{h_2^2+h_3^2}
    \end{pmatrix}, \\
    u^{(3)} =& \frac{1}{\sqrt{h_4^2+h_5^2}}\begin{pmatrix}
        h_4 & h_5 \\
        h_5 & -h_4
    \end{pmatrix}. 
\end{split}
\end{equation}
So far the circuit \eqref{eq:bogo_as_brickwall} is a map acting on mode operators (or equivalently, on single particle wavefunctions). We need to promote it to a circuit acting on the whole (truncated) Fock space. This is conceptually straightforward even if practically tedious. To this end, consider a number state of two modes $a,b$ and act with a $2\times 2$ unitary $u$ at the mode level:
\begin{align}
  \ket{n_a,n_b} &:= \frac{\ha^{\dagger n_a} \hb^{\dagger n_b}}{\sqrt{n_a!n_b!}}\ket{0}\\ 
                &\underset{u^{(i)}}{\longrightarrow} \frac{(u_{11} a + u_{21} b)^{n_a} (u_{12} a + u_{22} b)^{n_b}
                }{\sqrt{n_a!n_b!}}\ket{0} \label{eq:mode_transformed_factorized}\\ 
&:= U^{(i)} \ket{n_a,n_b}\, .
\end{align}
where $u_{nm}$ are the matrix entries of $u$.
This defines a unitary operator $U^{(i)}$ acting directly on Fock space, and that has the same effect as applying $u^{(i)}$ at the mode level. By expanding the right-hand side of \eqref{eq:mode_transformed_factorized}, we find the rows of $U^{(i)}$ one by one:
\begin{widetext}
\begin{eqnarray}
    \bra{\tilde{n}_L,\tilde{n}_R} U^{(i)} \ket{n_L,n_R} &=& \delta_{\tilde{n}_L+\tilde{n}_R,n_L+n_R} \sqrt{\frac{\tilde{n}_L!\tilde{n}_R!}{n_L! n_R!}}\sum_{k_L=\max(0,\tilde{n}_R-n_R)}^{\min(n_L,\tilde{n}_R)} C_{k_L,\tilde{n}_R-k_L}^{n_L,n_R},
\end{eqnarray}
where the coefficient $C_{k_L,\tilde{n}_R-k_L}^{n_L,n_R}$ is given by
\begin{equation}
    C_{k_L,\tilde{n}_R-k_L}^{n_L,n_R} = \binom{n_L}{k_L}\binom{n_R}{\tilde{n}_R-k_L} \left(u^{(i)}_{11}\right)^{n_L-k_L} \left(u^{(i)}_{12}\right)^{n_R-\tilde{n}_R+k_L} \left(u^{(i)}_{21}\right)^{k_L} \left(u^{(i)}_{22}\right)^{\tilde{n}_R-k_L}.
\end{equation}
\end{widetext}
We then get the brick-wall quantum circuit advertised by promoting $u^{(i)}$ to $U^{(i)}$ in \eqref{eq:bogo_as_brickwall}.

At the end of this procedure, we have simply rewritten the state $\ket{A,r}\in\mathcal{H}^r$ as a tensor network adapted to the tensor product structure of $\mathcal{H}^{r+1}$:
\begin{equation}\label{eq:iwt_equality}
  \vcenter{\hbox{\begin{tikzpicture}[line width=0.6pt,
    Abox/.style={rectangle, rounded corners, minimum width=0.6cm, minimum height=0.6cm, draw, fill=white},
    U1box/.style={rectangle, minimum width=1.6cm, minimum height=0.6cm, draw, fill=green!20},
    U2box/.style={rectangle, minimum width=1.6cm, minimum height=0.6cm, draw, fill=blue!20},
    U3box/.style={rectangle, minimum width=1.6cm, minimum height=0.6cm, draw, fill=red!20},
]

\begin{scope}
    \draw (-0.45,1.6) -- (4.45,1.6);
    \foreach \x in {0.8,3.2}{ \node[Abox] at (\x,1.6) {$A$}; }
    \foreach \x in {0.8,3.2}{ \draw (\x,1.3) -- (\x,0.6); }
    \node[font=\large] at (2.0,-3.1) {$\in\;\mathcal{H}^{r}$};
\end{scope}

\node at (5.1,1.6) {\Large $\cong$};

\begin{scope}[xshift=6.0cm]
    \draw (-0.45,1.6) -- (4.45,1.6);
    \foreach \x in {0.8,3.2}{ \node[Abox] at (\x,1.6) {$A$}; }
    \foreach \xc/\xs/\xv in {0.8/0.2/1.4, 3.2/2.6/3.8}{
        \draw (\xc,1.3) -- (\xc,1.0);
        \draw (\xc,1.0) -- (\xs,0.3);
        \node at (\xv,0.6) {$\ket{0}$};
    }
    \foreach \x in {0.2,2.6}{ \draw (\x,0) -- (\x,0.3); }
    \foreach \x in {1.4,3.8}{ \draw (\x,0) -- (\x,0.4); }
    \foreach \x in {0.8,3.2}{ \node[U1box] at (\x,-0.3) {$U^{(1)}$}; }
    \foreach \x in {0.2,1.4,2.6,3.8}{ \draw (\x,-0.6) -- (\x,-0.9); }
    \node[U2box] at (2.0,-1.2) {$U^{(2)}$};
    \draw[fill=blue!20] (-0.45,-0.9) -- (0.4,-0.9) -- (0.4,-1.5) -- (-0.45,-1.5);
    \draw[fill=blue!20] (4.45,-0.9) -- (3.6,-0.9) -- (3.6,-1.5) -- (4.45,-1.5);
    \foreach \x in {0.2,1.4,2.6,3.8}{ \draw (\x,-1.5) -- (\x,-1.8); }
    \foreach \x in {0.8,3.2}{ \node[U3box] at (\x,-2.1) {$U^{(3)}$}; }
    \foreach \x in {0.2,1.4,2.6,3.8}{ \draw (\x,-2.4) -- (\x,-2.7); }
    \node[font=\large] at (2.0,-3.1) {$\in\;\mathcal{H}^{r+1}$};
\end{scope}

\end{tikzpicture}}}\, ,
\end{equation}
where we use the sign $\cong$ to mean that these different tensor networks, with legs living in different local Hilbert spaces, represent the same state in the global Hilbert space $\mathcal{H}^r\subset\mathcal{H}^{r+1}$.

As discussed in depth in~\cite{PhysRevLett.116.140403, PhysRevX.8.011003, PhysRevA.97.052314, Witteveen_Scholz_Swingle_Walter_2022, 10.21468/SciPostPhys.10.6.143}, a circuit that introduces wavelet modes and then performs a wavelet transform via two-site unitaries is an entanglement renormalization circuit. However, we do not stack several such layers at once, and thus depart from standard discussions connecting MERA and wavelets.

\subsection{Refinement map contraction via ITEBD}

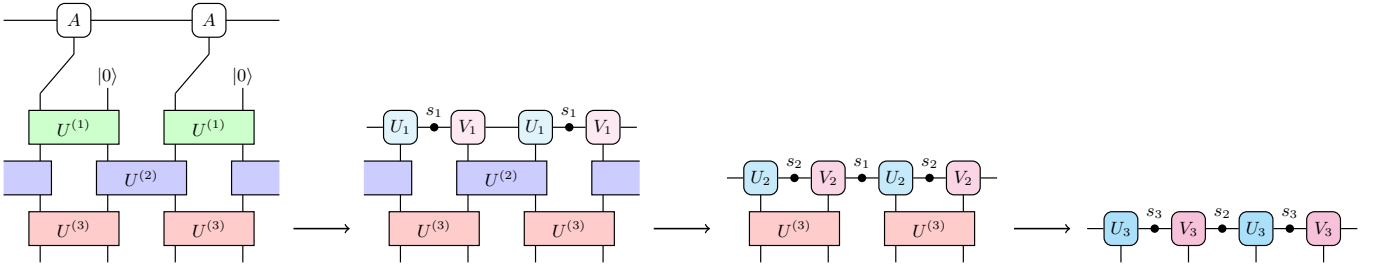
\begin{figure*}
    \centering
    \begin{tikzpicture}[line width=0.6pt,
    Abox/.style={rectangle, rounded corners, minimum width=0.6cm, minimum height=0.6cm, draw, fill=white},
    U1box/.style={rectangle, minimum width=1.6cm, minimum height=0.6cm, draw, fill=green!20},
    U2box/.style={rectangle, minimum width=1.6cm, minimum height=0.6cm, draw, fill=blue!20},
    U3box/.style={rectangle, minimum width=1.6cm, minimum height=0.6cm, draw, fill=red!20},
    dot/.style={circle, fill, inner sep=1.5pt},
]

\begin{scope}
    \draw (-0.45,1.6) -- (4.45,1.6);
    \foreach \x in {0.8,3.2}{ \node[Abox] at (\x,1.6) {$A$}; }
    \foreach \xc/\xs/\xv in {0.8/0.2/1.4, 3.2/2.6/3.8}{
        \draw (\xc,1.3) -- (\xc,1.0);
        \draw (\xc,1.0) -- (\xs,0.3);
        \node at (\xv,0.6) {$\ket{0}$};
    }
    \foreach \x in {0.2,2.6}{ \draw (\x,0) -- (\x,0.3); }
    \foreach \x in {1.4,3.8}{ \draw (\x,0) -- (\x,0.4); }
    \foreach \x in {0.8,3.2}{ \node[U1box] at (\x,-0.3) {$U^{(1)}$}; }
    \foreach \x in {0.2,1.4,2.6,3.8}{ \draw (\x,-0.6) -- (\x,-0.9); }
    \node[U2box] at (2.0,-1.2) {$U^{(2)}$};
    \draw[fill=blue!20] (-0.45,-0.9) -- (0.4,-0.9) -- (0.4,-1.5) -- (-0.45,-1.5);
    \draw[fill=blue!20] (4.45,-0.9) -- (3.6,-0.9) -- (3.6,-1.5) -- (4.45,-1.5);
    \foreach \x in {0.2,1.4,2.6,3.8}{ \draw (\x,-1.5) -- (\x,-1.8); }
    \foreach \x in {0.8,3.2}{ \node[U3box] at (\x,-2.1) {$U^{(3)}$}; }
    \foreach \x in {0.2,1.4,2.6,3.8}{ \draw (\x,-2.4) -- (\x,-2.7); }
\end{scope}

\draw[->, thick] (4.7,-2.1) -- (5.7,-2.1);

\begin{scope}[xshift=6.4cm]
    \node[Abox, fill=cyan!10]    (c1) at (0.2,-0.3) {$U_1$};
    \node[Abox, fill=magenta!10] (c2) at (1.4,-0.3) {$V_1$};
    \node[Abox, fill=cyan!10]    (c3) at (2.6,-0.3) {$U_1$};
    \node[Abox, fill=magenta!10] (c4) at (3.8,-0.3) {$V_1$};
    \draw (-0.4,-0.3) -- (c1.west); \draw (c1.east) -- (c2.west); \draw (c2.east) -- (c3.west);
    \draw (c3.east) -- (c4.west); \draw (c4.east) -- (4.4,-0.3);
    \node[dot] at (0.8,-0.3) {}; \node at (0.8,-0.03) {$s_1$};
    \node[dot] at (3.2,-0.3) {}; \node at (3.2,-0.03) {$s_1$};
    \foreach \x in {0.2,1.4,2.6,3.8}{ \draw (\x,-0.6) -- (\x,-0.9); }
    \node[U2box] at (2.0,-1.2) {$U^{(2)}$};
    \draw[fill=blue!20] (-0.45,-0.9) -- (0.4,-0.9) -- (0.4,-1.5) -- (-0.45,-1.5);
    \draw[fill=blue!20] (4.45,-0.9) -- (3.6,-0.9) -- (3.6,-1.5) -- (4.45,-1.5);
    \foreach \x in {0.2,1.4,2.6,3.8}{ \draw (\x,-1.5) -- (\x,-1.8); }
    \foreach \x in {0.8,3.2}{ \node[U3box] at (\x,-2.1) {$U^{(3)}$}; }
    \foreach \x in {0.2,1.4,2.6,3.8}{ \draw (\x,-2.4) -- (\x,-2.7); }
\end{scope}

\draw[->, thick] (11.1,-2.1) -- (12.1,-2.1);

\begin{scope}[xshift=12.8cm]
    \node[Abox, fill=cyan!20]    (d1) at (0.2,-1.2) {$U_2$};
    \node[Abox, fill=magenta!20] (d2) at (1.4,-1.2) {$V_2$};
    \node[Abox, fill=cyan!20]    (d3) at (2.6,-1.2) {$U_2$};
    \node[Abox, fill=magenta!20] (d4) at (3.8,-1.2) {$V_2$};
    \draw (-0.4,-1.2) -- (d1.west); \draw (d1.east) -- (d2.west); \draw (d2.east) -- (d3.west);
    \draw (d3.east) -- (d4.west); \draw (d4.east) -- (4.4,-1.2);
    \node[dot] at (0.8,-1.2) {}; \node at (0.8,-0.93) {$s_2$};
    \node[dot] at (2.0,-1.2) {}; \node at (2.0,-0.93) {$s_1$};
    \node[dot] at (3.2,-1.2) {}; \node at (3.2,-0.93) {$s_2$};
    \foreach \x in {0.2,1.4,2.6,3.8}{ \draw (\x,-1.5) -- (\x,-1.8); }
    \foreach \x in {0.8,3.2}{ \node[U3box] at (\x,-2.1) {$U^{(3)}$}; }
    \foreach \x in {0.2,1.4,2.6,3.8}{ \draw (\x,-2.4) -- (\x,-2.7); }
\end{scope}

\draw[->, thick] (17.5,-2.1) -- (18.5,-2.1);

\begin{scope}[xshift=19.2cm]
    \node[Abox, fill=cyan!30]    (f1) at (0.2,-2.1) {$U_3$};
    \node[Abox, fill=magenta!30] (f2) at (1.4,-2.1) {$V_3$};
    \node[Abox, fill=cyan!30]    (f3) at (2.6,-2.1) {$U_3$};
    \node[Abox, fill=magenta!30] (f4) at (3.8,-2.1) {$V_3$};
    \draw (-0.4,-2.1) -- (f1.west); \draw (f1.east) -- (f2.west); \draw (f2.east) -- (f3.west);
    \draw (f3.east) -- (f4.west); \draw (f4.east) -- (4.4,-2.1);
    \node[dot] at (0.8,-2.1) {}; \node at (0.8,-1.83) {$s_3$};
    \node[dot] at (2.0,-2.1) {}; \node at (2.0,-1.83) {$s_2$};
    \node[dot] at (3.2,-2.1) {}; \node at (3.2,-1.83) {$s_3$};
    \foreach \x in {0.2,1.4,2.6,3.8}{ \draw (\x,-2.4) -- (\x,-2.7); }
\end{scope}

\end{tikzpicture}
    \caption{Contraction of the original MPS with three layers of the IWT brick-wall circuit into an MPS with a larger unit cell using ITEBD, which corresponds to step (ii) of our refinement algorithm.}
    \label{fig:tebd}
\end{figure*}

We now need to contract the right-hand side of \eqref{eq:iwt_equality} to bring it back to MPS form. We do so with $3$ steps of the ITEBD algorithm, contracting the brick-wall layers one at a time as summarized in Fig.~\ref{fig:tebd}.

For the first layer, at each site we contract the MPS tensor $A$, the wavelet state, and $U^{(1)}$. We carry a singular value decomposition (SVD) on the resulting tensor
\begin{center}
    \begin{tikzpicture}[line width=0.6pt]

\node[rectangle, rounded corners,minimum width=0.6cm,minimum height=0.6cm, draw] (A1) at (0.8, 1.6) {$A$};

\draw (0, 1.6) -- (A1.west);
\draw (A1.east) -- (1.6,1.6);

\draw (0.8,1.0) -- (0.2,0.3);

\draw (A1.south) -- (0.8, 1.0);

\node[rectangle,minimum width=1.6cm,minimum height=0.6cm, draw,fill=green!20] (a11) at (0.8,-0.3) {$U^{(1)}$};

\draw (0.2, 0) -- (0.2, 0.3);
\draw (1.4, 0) -- (1.4, 0.4);
\draw (0.2, -0.6) -- (0.2, -0.9);
\draw (1.4, -0.6) -- (1.4, -0.9);

\draw[->, thick] (2.2,0.3) -- (3.2, 0.3);
\draw (2.7,0.6) node {Contract};
\draw[->, thick] (5.9,0.3) -- (6.9, 0.3);
\draw (6.4,0.6) node {SVD};
\draw (1.4,0.6) node {$\ket{0}$};

\begin{scope}[xshift=3.8cm, yshift=-1.05cm]
    \node[rectangle, rounded corners, minimum width=1cm,minimum height=0.8cm, fill=gray!20, draw] (A1) at (0.8, 1.6) {};

\draw (0, 1.6) -- (A1.west);
\draw (A1.east) -- (1.6,1.6);

\draw (0.45,1.2) -- (.45,.9);
\draw (1.15,1.2) -- (1.15,.9);

\draw[blue!70, thick, dashed] (.8,2.5) -- (.8, .7);
\end{scope}

\begin{scope}[xshift=7.4cm]
    \node[rectangle, rounded corners,minimum width=0.6cm,minimum height=0.6cm, fill=cyan!10, draw] (U1) at (0.8,0.6) {$U_1$};
    \node[rectangle, rounded corners,minimum width=0.6cm,minimum height=0.6cm, fill=magenta!10, draw] (V1) at (2.8, 0.6) {$V_1$};

    \draw (0.0, 0.6) -- (U1.west);
    \draw (U1.east) -- (V1.west);
    \draw (V1.east) -- (3.6,0.6);
    \draw (U1.south) -- (0.8, -0.1);
    \draw (V1.south) -- (2.8, -0.1);

    \node (s1) at (1.8,0.6) [circle,fill,inner sep=1.5pt]{};
    \draw (1.8,0.85) node {$s_1$};
\end{scope}

\end{tikzpicture}
\end{center}
to get an infinite two-site MPS with tensors $U_1$ and $s_1V_1$. The second layer is similarly contracted and decomposed as
\begin{center}
    \begin{tikzpicture}[line width=0.6pt]

\node[rectangle, rounded corners,minimum width=0.6cm,minimum height=0.6cm, fill=magenta!10, draw] (V1) at (1.2,1) {$V_1$};
\node[rectangle, rounded corners,minimum width=0.6cm,minimum height=0.6cm, fill=cyan!10, draw] (U1) at (2.5, 1) {$U_1$};

\draw (0.0, 1) -- (V1.west);
\draw (V1.east) -- (U1.west);
\draw (U1.east) -- (3.7,1);
\draw (V1.south) -- (1.2, 0.3);
\draw (U1.south) -- (2.5, 0.3);

\node (s1) at (0.45,1) [circle,fill,inner sep=1.5pt]{};
\draw (0.45,1.25) node {$s_1$};

\node at (3.25,1) [circle,fill,inner sep=1.5pt]{};
\draw (3.25,1.25) node {$s_1$};

\node[rectangle,minimum width=1.6cm,minimum height=0.6cm, draw,fill=blue!20] (a2) at (1.85,0) {$U^{(2)}$};
\draw (1.2,-0.3) -- (1.2, -0.7);
\draw (2.5,-0.3) -- (2.5, -0.7);

\draw[->, thick] (4.6,0.3) -- (5.6, 0.3);
\draw (5.1,0.6) node {SVD};

\begin{scope}[xshift=6.5cm]
    \node[rectangle, rounded corners,minimum width=0.6cm,minimum height=0.6cm, fill=cyan!20, draw] (U2) at (1.2,0.6) {$U_2$};
    \node[rectangle, rounded corners,minimum width=0.6cm,minimum height=0.6cm, fill=magenta!20, draw] (V2) at (3.2, 0.6) {$V_2$};

    \draw (0.0, 0.6) -- (U2.west);
    \draw (U2.east) -- (V2.west);
    \draw (V2.east) -- (4.4,0.6);
    \draw (U2.south) -- (1.2, -0.1);
    \draw (V2.south) -- (3.2, -0.1);

    \node at (2.2,0.6) [circle,fill,inner sep=1.5pt]{};
    \draw (2.2,0.85) node {$s_2$};

    \node at (0.45,0.6) [circle,fill,inner sep=1.5pt]{};
    \draw (0.45,0.85) node {$s_1$};

    \node at (3.95,0.6) [circle,fill,inner sep=1.5pt]{};
    \draw (3.95,0.85) node {$s_1$};
\end{scope}

\end{tikzpicture}
\end{center}
and the process for the final layer follows by analogy. For the last two layers, the contraction includes the singular values $s_{i-1}$, and the SVD yields $\tilde{U}_is_i\tilde{V}_i$ with the $s_{i-1}$ singular values absorbed into $\tilde{U}_i,\tilde{V}_i$. To obtain the final diagram in the figure above, one takes inverses of the singular values such that $U_i = s_{i-1}^{-1}\tilde{U}_i$ and $V_i = \tilde{V}_i s_{i-1}^{-1}$.

The resulting state is an MPS with a 2-site unit cell at resolution $r+1$, with tensors $A$ and $B$ (for example, we take $A= s_3 U_3$ and $B=s_2 V_3$), and much larger bond dimension. In practice, we truncate the bond dimension back to $\chi$ at the end, and observe that this only minimally impacts the energy density of the state.

\subsection{Approximate projection onto translation invariant MPS}
If we iterate the steps (i) and (ii) that we presented before $M$ times, the unit cell will have size $2^M$. Hence, before refining further, we need to get back to an MPS with a $1$-site unit cell. We should do so in a way that does not degrade the energy density. Fortunately, we expect the ground state to be translation invariant, and thus we expect the two-site MPS we obtain to be approximately translation invariant.

If the state were exactly translation invariant, then it would imply that there exists some unitary $U$ for which $U^\dag A = BU = C$, explicitly restoring translation invariance. Note that translation invariance would automatically be restored by subsequent optimization (\textit{e.g.} with VUMPS). However, by approximately projecting onto translation invariant MPS directly, we get two benefits: first, VUMPS is cheaper when run on single-site MPS, and second, we see that we have a discrete freedom in the projection that allows us to reach a substantially lower energy.

To find $U$, we first put $A,B$ in right-orthonormal gauge, to get $A_R,B_R$, respectively. We then construct the following transfer matrix \cite{forum}:
\begin{center}
    \begin{tikzpicture}[line width=0.6pt]

\node[rectangle, rounded corners,minimum width=0.6cm,minimum height=0.6cm, fill=cyan!40, inner sep=0cm, draw] (A) at (1,0.6) {$A_R$};
\node[rectangle, rounded corners,minimum width=0.6cm,minimum height=0.6cm, fill=magenta!40, inner sep=0cm, draw] (B) at (2.4, 0.6) {$B_R$};

\node[rectangle, rounded corners,minimum width=0.6cm,minimum height=0.6cm, fill=magenta!40, inner sep=0cm, draw] (Bb) at (1,-0.6) {$\bar{B}_R$};
\node[rectangle, rounded corners,minimum width=0.6cm,minimum height=0.6cm, fill=cyan!40, inner sep=0cm, draw] (Ab) at (2.4, -0.6) {$\bar{A}_R$};

\draw (A.south) -- (Bb.north);
\draw (B.south) -- (Ab.north);
\draw (A.east) -- (B.west);
\draw (Bb.east) -- (Ab.west);
    
\draw (0,0.6) -- (A.west);
\draw (0,-0.6) -- (Bb.west);
\draw (B.east) -- (3.4,0.6);
\draw (Ab.east) -- (3.4,-0.6);

\draw (4.0,0) node {$=$};
\draw (4.0,-2.7) node {$=$};

\begin{scope}[xshift=4.6cm]
    \node[circle, minimum width=0.6cm, inner sep=0cm, draw] (U1) at (0.7,0.6) {$U$};
    \node[circle, minimum width=0.6cm, inner sep=0cm, draw] (Ud1) at (1.6,0.6) {$U^\dag$};
    \node[rectangle, rounded corners,minimum width=0.6cm,minimum height=0.6cm, fill=cyan!40, inner sep=0cm, draw] (A) at (2.5,0.6) {$A_R$};
    \node[rectangle, rounded corners,minimum width=0.6cm,minimum height=0.6cm, fill=magenta!40, inner sep=0cm, draw] (B) at (5.2, 0.6) {$B_R$};
    \node[circle, minimum width=0.6cm, inner sep=0cm, draw] (U2) at (6.1,0.6) {$U$};
    \node[circle, minimum width=0.6cm, inner sep=0cm, draw] (Ud2) at (7,0.6) {$U^\dag$};

    \node[rectangle, rounded corners,minimum width=0.6cm,minimum height=0.6cm, fill=magenta!40, inner sep=0cm, draw] (Bb) at (2.5,-0.6) {$\bar{B}_R$};
    \node[circle, minimum width=0.6cm, inner sep=0cm, draw] (Ub) at (3.4,-0.6) {$\bar{U}$};
    \node[circle, minimum width=0.6cm, inner sep=0cm, draw] (Ubd) at (4.3,-0.6) {$\bar{U}^\dag$};
    \node[rectangle, rounded corners,minimum width=0.6cm,minimum height=0.6cm, fill=cyan!40, inner sep=0cm, draw] (Ab) at (5.2, -0.6) {$\bar{A}_R$};
    
    \draw (A.south) -- (Bb.north);
    \draw (B.south) -- (Ab.north);

    \draw (0,0.6) -- (U1.west);
    \draw (U1.east) -- (Ud1.west);
    \draw (Ud1.east) -- (A.west);
    \draw (A.east) -- (B.west);
    \draw (B.east) -- (U2.west);
    \draw (U2.east) -- (Ud2.west);
    \draw (Ud2.east) -- (7.7,0.6);
    
    \draw (0,-0.6) -- (Bb.west);
    \draw (Bb.east) -- (Ub.west);
    \draw (Ub.east) -- (Ubd.west);
    \draw (Ubd.east) -- (Ab.west);
    \draw (Ab.east) -- (7.7,-0.6);
\end{scope}

\begin{scope}[xshift=4.6cm, yshift=-2.7cm]
    \node[circle, minimum width=0.6cm, inner sep=0cm, draw] (U) at (0.7,0.6) {$U$};
    \node[rectangle, rounded corners,minimum width=0.6cm,minimum height=0.6cm, fill=violet!30, inner sep=0cm, draw] (C1) at (1.6,0.6) {$C_R$};
    \node[rectangle, rounded corners,minimum width=0.6cm,minimum height=0.6cm, fill=violet!30, inner sep=0cm, draw] (C2) at (3, 0.6) {$C_R$};
    \node[circle, minimum width=0.6cm, inner sep=0cm, draw] (Ud) at (3.9,0.6) {$U^\dag$};

    \node[rectangle, rounded corners,minimum width=0.6cm,minimum height=0.6cm, fill=violet!30, inner sep=0cm, draw] (Cb1) at (1.6,-0.6) {$\bar{C}_R$};
    \node[rectangle, rounded corners,minimum width=0.6cm,minimum height=0.6cm, fill=violet!30, inner sep=0cm, draw] (Cb2) at (3, -0.6) {$\bar{C}_R$};
    
    \draw (C1.south) -- (Cb1.north);
    \draw (C2.south) -- (Cb2.north);

    \draw (0,0.6) -- (U.west);
    \draw (U.east) -- (C1.west);
    \draw (C1.east) -- (C2.west);
    \draw (C2.east) -- (Ud.west);
    \draw (Ud.east) -- (4.6,0.6);

    \draw (0,-0.6) -- (Cb1.west);
    \draw (Cb1.east) -- (Cb2.west);
    \draw (Cb2.east) -- (4.6,-0.6);
\end{scope}
\end{tikzpicture}
\end{center}
$C_R=U^\dag A_R = B_R U$ which is possible only if the state is exactly translation invariant (in which case, making translation invariance explicit, and finding $C$, is a pure gauge choice). Importantly, $U$ is a fixed point of this transfer matrix with eigenvalue 1:
\begin{center}
    \begin{tikzpicture}[line width=0.6pt]

    \node[circle, minimum width=0.6cm, inner sep=0cm, draw] (U) at (0.7,0.6) {$U$};
    \node[rectangle, rounded corners,minimum width=0.6cm,minimum height=0.6cm, fill=violet!30, inner sep=0cm, draw] (C1) at (1.6,0.6) {$C_R$};
    \node[rectangle, rounded corners,minimum width=0.6cm,minimum height=0.6cm, fill=violet!30, inner sep=0cm, draw] (C2) at (3, 0.6) {$C_R$};
    \node[circle, minimum width=0.6cm, inner sep=0cm, draw] (Ud) at (3.9,0.6) {$U^\dag$};

    \node[rectangle, rounded corners,minimum width=0.6cm,minimum height=0.6cm, fill=violet!30, inner sep=0cm, draw] (Cb1) at (1.6,-0.6) {$\bar{C}_R$};
    \node[rectangle, rounded corners,minimum width=0.6cm,minimum height=0.6cm, fill=violet!30, inner sep=0cm, draw] (Cb2) at (3, -0.6) {$\bar{C}_R$};
    
    \draw (C1.south) -- (Cb1.north);
    \draw (C2.south) -- (Cb2.north);

    \draw (0,0.6) -- (U.west);
    \draw (U.east) -- (C1.west);
    \draw (C1.east) -- (C2.west);
    \draw (C2.east) -- (Ud.west);
    \draw (Ud.east) -- (4.4,0.6);

    \draw (0,-0.6) -- (Cb1.west);
    \draw (Cb1.east) -- (Cb2.west);
    \draw (Cb2.east) -- (4.4,-0.6);

    \node[circle, minimum width=0.6cm, inner sep=0cm, draw=red!50, text=red!50] (Ueig) at (4.9,0.6) {$U$};
    \draw[red] (4.4,0.6) -- (4.6,0.6);
    \draw[red, rounded corners=6pt] (Ueig.east) -- (5.5,0.6) -- (5.5,-0.6) -- (4.4,-0.6);

    \draw (6.1,0) node {$=$};

\begin{scope}[xshift=6.7cm]
    \node[circle, minimum width=0.6cm, inner sep=0cm, draw] (Uf) at (0.7,0.6) {$U$};

    \draw (0,0.6) -- (Uf.west);
    \draw[rounded corners=6pt] (Uf.east) -- (1.4,0.6) -- (1.4,-0.6) -- (0,-0.6);
    
\end{scope}

\end{tikzpicture}
\end{center}
This allows one to find $U$, then $C_R$, and thus approximate the refined two-site MPS as a single-site state $\ket{C_R, r+1}$. We say \emph{approximate}, because in practice, $U^\dag A_R$ and $B_R U$ are not exactly equal, so we are free to choose whichever one gives a state with lower energy density. Quite surprisingly, we observe that one choice systematically \emph{lowers} the energy compared to the original two-site state, almost bringing it to the energy of the fully optimized state at resolution $r+1$!

This is the final step of the refinement algorithm. We now have a process which starts at an initial single-site MPS $\ket{A, r}$ at resolution $r$ with energy density $e_0(r)$, and outputs a single-site state $\ket{\tilde A, r+1}$ with lower or equal energy density.

\subsection{Numerical results}\label{sec:fixed_vs_multi}

\begin{figure}
    \centering
    \includegraphics[width=\linewidth]{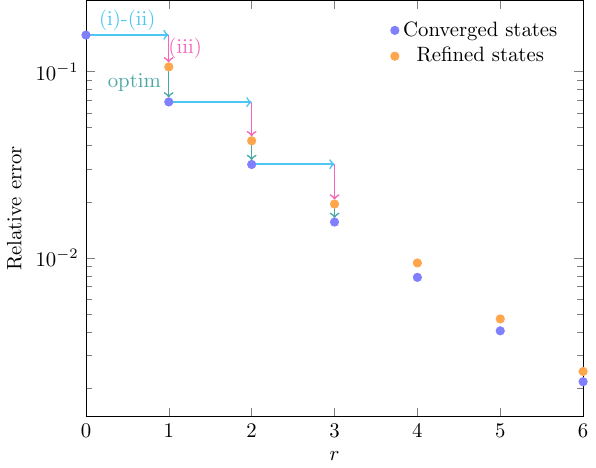}
    \caption{Comparing the output of the refinement algorithm (orange markers) before subsequent optimization to the converged energy densities (blue markers) at each resolution, with $\chi=16,\mu=1,c=8,N=6$. From a converged ground state, we act with steps (i)-(ii) of the refinement algorithm (blue arrows), which keeps the energy the same up to small errors. By imposing translation invariance (pink arrows, step (iii) of the refinement algorithm), the error in energy density decreases significantly. Then we optimize with a combination of VUMPS and gradient descent (green arrows) to get the converged state.}
    \label{fig:staircase}
\end{figure}

We first look at the behavior of the energy density at each of the scale refinement steps in Fig. \ref{fig:staircase}. As expected, the energy stays constant as we contract the brick-wall circuit and our resolution $r$ MPS to get a two-site MPS at resolution $r+1$. Then, quite surprisingly, approximately projecting onto a one-site MPS drastically lowers the energy density -- it even gets closer and closer to its post-optimization value at $r+1$ when $r$ increases. Adding a few steps of an optimization algorithm at the end of the refinement gets us to the best possible energy at $r+1$.

This iterative refinement approach is in fact the only way we manage to find approximate ground states in a reasonable amount of time for $r\geq 6$. 

\begin{figure}
    \centering
    \includegraphics[width=\linewidth]{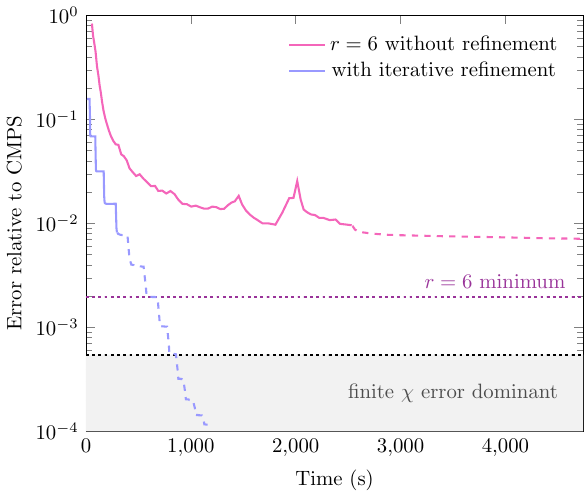}
    \caption{Comparing optimization routines with and without use of the refinement algorithm, for $\mu=1,c=8, \chi=12$. Without refinement, we start at a random $r=6$ state, iterate with VUMPS until it becomes unstable, then optimize with LBFGS (dashed curve). With the refinement algorithm, we start at a random $r=0$ state, optimize until convergence, and then refine to $r+1$. We repeat this process until we reach $r=12$, where the error in the energy density is fully dominated by finite bond dimension effects. For illustrative purposes we start at 10 VUMPS iterations for both methods. The time reported is for a MacBook Pro M3.} 
    \label{fig:timing}
\end{figure}

We consider that we have sufficiently refined in scale when the energy density error is dominated by the finite $\chi$ error, and not the scale error. On this toy model, we can distinguish the two cleanly by computing the CMPS energy density: when $e_0^\text{wMPS}(r) - e_0^\text{CMPS} \ll e_0^\text{CMPS} - e_0^\text{exact}$ we are at a scale that is small enough. In Fig. \ref{fig:timing}, we show that without iterative refinement, optimizing at $r=6$ directly, using VUMPS (on the edge of stability) followed by Grassmann gradient descent, the optimization does not converge (staying far from what iterative refinement reaches at the same $r$) and stays quite far from the threshold where the finite bond dimension error dominates. In contrast, when we alternate refinements and Grassmann gradient descent, we can reach much higher $r$ and go to a regime where scale error is negligible.

\section{Discussion}
We have presented a new method to solve interacting continuum models with Daubechies scaling functions and MPS. The main advantage of the approach is that it is strictly variational: it gives certified upper bounds on the energy, and a clean way to compare the quality of different ground-state approximations. Using an iterative refinement scheme, which exploits multi-resolution analysis, it is possible to get accurate results on the example of the Lieb-Liniger model. In that case, for moderate bond dimension, the residual error is dominated by finite entanglement effects, and not the way one deals with the continuum.

Importantly, this method is compatible with the entirety of the discrete tensor network toolbox, as it is already implemented in standard libraries. While we demonstrated the power of the method on a translation invariant ground state problem with uniform MPS -- because it makes comparison with exact results easier -- it can be used on problems with open boundary conditions, for simulations of real-time dynamics, or to compute excitation spectra, simply by plugging in the corresponding MPS routine. This flexibility is the main advantage of wMPS over dedicated continuous algorithms like CMPS. The latter may however keep an edge in the thermodynamic limit when high accuracy and large bond dimensions are required.

The link between wMPS and CMPS remains to be fully elucidated. We did observe that a wMPS at $r\rightarrow + \infty$ gives the same energy density as a CMPS of identical bond dimension. This naturally raises the question of how well one can discretize a CMPS into wMPS, which would allow one to conveniently change representation when needed. Since the theory of CMPS can be mapped to the theory of continuous measurement~\cite{Osborne2010Holographic}, this problem is related to optimal ways of binning time-continuous measurements into discrete records~\cite{guilmin2025timeaveragedcontinuousquantummeasurement,wonglakhon2026quantumtrajectoriestimebinneddata}, which is particularly relevant for the manipulation of superconducting qubits.

A natural next step would be to consider broader classes of wavelets beyond the Daubechies family. For example, relaxing orthonormality, with biorthogonal wavelets (as used \textit{e.g.} in~\cite{10.21468/SciPostPhys.10.6.143}), would allow the use of more symmetric and regular wavelets with the same support. More generally, it would be interesting to understand what criterion one should optimize to get the ``best'' function with fixed support for typical variational problems.

Another extension would be to consider relativistic QFT, where extra divergences (related to short-distance scale invariance) bring additional difficulties. In this context, relativistic CMPS (RCMPS) solve the problem, and are truly variational, but at the cost of extra non-locality, which translates into a higher numerical cost and the need for an \textit{ad hoc} optimizer. If it is possible to use standard MPS algorithms while remaining variational in the relativistic case, this would be particularly useful.

Finally, there is in principle no obstruction to generalizing our wMPS to higher dimensions, with wavelet Projected Entangled Pair States (wPEPS). In practice, the bond dimension of the MPO in one dimension is already large, and so we anticipate it will be expensive to perform a numerical analysis with a wPEPS. In this context, however, there is less competition with continuous methods. In particular, despite some recent progress with continuous tensor network states~\cite{jennings2015,PhysRevX.9.021040,shachar2022,rigobello2026gaussiancontinuoustensornetwork,roose2026continuumlimitgaugedtensor}, which can now be contracted in some non-trivial cases~\cite{tiwana2026}, there is no functional tensor network algorithm to treat the 2d Bose gas directly in the continuum. 

\begin{acknowledgements}
  We thank Claudia Frugiuele, Jutho Haegeman, Hector Hutin, and Slava Rychkov for particularly enlightening discussions on the subject of this paper. This project has received funding from the European Research Council (ERC) under the European Union's Horizon 2020 research and innovation programme (grant agreement No.~884762), and also from the ERC QFT.zip project (grant agreement No.~101040260).
\end{acknowledgements}

\bibliography{draft}

\appendix*

\section{Evaluating scaling function integrals}\label{app:ints}
We detail the calculation of $K_{a}$ in eq.~\eqref{eq:K} and $\Gamma^4_{a,b,c}$ in eq.~\eqref{eq:Gamma4}, for any $N$. As described in~\cite{PhysRevD.101.096004}, the key to evaluating these coefficients without performing any integration is the fixed-point equation~\eqref{eq:RGs}. Starting with the kinetic coefficient, with $-N+2 \leq a \leq N-2$, we use the fixed-point equation to find
\begin{eqnarray}
    K_a &=& 2 \sum_{m,n=0}^{N-1} h_m h_n \int \upd x\,\partial_x s(2x -m) \partial_x s(2x-2a -n) \nonumber \\
    &=& 4 \sum_{m,n=0}^{N-1} h_m h_n \int \upd x\,\partial_x s(x) \partial_x s(x-2a -n + m) \nonumber \\
    &=& 4 \sum_{m,n=0}^{N-1} h_m h_n K_{2a+n-m} \nonumber \\
    &=& 4 \sum_{n'=-N+2}^{N-2} \sum_{m=0}^{N-1} h_m h_{n'+m-2a} K_{n'} \nonumber \\
    &:=& \sum_{n'=-N+2}^{N-2} \mathcal{A}_{an'} K_{n'}
\end{eqnarray}
where we defined $ \mathcal{A}_{an'}  = 4 \sum_{m=0}^{N-1} h_m h_{n'+m-2a}$. This then makes it clear that $K_a$ is an eigenvector of $\mathcal{A}$ with eigenvalue 1. It is normalized by the condition
\begin{equation}
    \sum_{a=-N+2}^{N-2} a^2K_a=-2
\end{equation}
which is derived in \cite{PhysRevD.101.096004} from the property that $s_n(x)$ can locally represent polynomials up to degree $N/2-1$.

The four-point coefficient in eq.~\eqref{eq:Gamma4} can be calculated in a similar fashion. Via the fixed-point equation~\eqref{eq:RGs}, $\Gamma^4$ becomes
\begin{widetext}
\begin{eqnarray}
    \Gamma^4_{a,b,c} &=& 4 \sum_{m,n,l,k=0}^{N-1} h_m h_n h_l h_k \int \upd x\,s(2x-m) s(2x - 2a -n) s(2x - 2b -l) s(2x - 2c -k) \nonumber \\
    &=& 2 \sum_{m,n,l,k=0}^{N-1} h_m h_n h_l h_k \int \upd x\,s(x) s(x - 2a -n +m) s(x - 2b -l+m) s(x - 2c -k+m) \nonumber \\
    &=& 2 \sum_{m,n,l,k=0}^{N-1} h_m h_n h_l h_k \Gamma^4_{2a+n-m,\, 2b+l-m,\, 2c+k-m} \nonumber \\
    &=& 2 \sum_{n',l',k'=-N+2}^{N-2} \sum_{m=0}^{N-1} h_m h_{n'+m-2a} h_{l'+m-2b} h_{k'+m-2c} \Gamma^4_{n',l',k'} \nonumber \\
    &:=& \sum_{n',l',k'=-N+2}^{N-2} \mathcal{B}_{n'l'k'}^{abc} \Gamma^4_{n',l',k'},
\end{eqnarray}
\end{widetext}
where we defined 
\begin{equation}
    \mathcal{B}_{n'l'k'}^{abc}= 2 \sum_{m=0}^{N-1} h_m h_{n'+m-2a} h_{l'+m-2b} h_{k'+m-2c}.
\end{equation}
Then $\Gamma^4_{a,b,c}$ is a fixed point of the tensor $\mathcal{B}$, with eigenvalue 1. Again, by the polynomial representation property, we normalize $\Gamma^4$ using
\begin{equation}
    \sum_{a,b=-N+2}^{N-2} \Gamma^4_{a,b,c} =\delta_{c0}.
\end{equation}
In this way, we find the kinetic and interaction coefficients for $N=6,8$.

\end{document}